\documentclass[%
 reprint,
superscriptaddress,
 amsmath,amssymb,
 aps,
]{revtex4-2}

\usepackage{graphicx}
\usepackage{dcolumn}
\usepackage{bm}
\usepackage{subfigure}
\usepackage{amsmath}
\usepackage{amssymb}

\usepackage{hyperref}
\hypersetup{
    colorlinks=true
}

\begin{document}

\preprint{APS/123-QED}

\title{Simulation of radio signals from cosmic-ray cascades in air and ice as observed by in-ice Askaryan radio detectors}

\author{Simon De Kockere}%
\email{simon.de.kockere@vub.be}
\affiliation{Vrije Universiteit Brussel, Dienst ELEM, IIHE, 1050 Brussels, Belgium}

\author{Dieder Van den Broeck}
\email{dieder.jan.van.den.broeck@vub.be}
\affiliation{Vrije Universiteit Brussel, Astrophysical Institute, IIHE, 1050 Brussels, Belgium}

\author{Uzair Abdul Latif}
\affiliation{Vrije Universiteit Brussel, Dienst ELEM, IIHE, 1050 Brussels, Belgium}

\author{Krijn D. de Vries}
\email{krijn.de.vries@vub.be}
\affiliation{Vrije Universiteit Brussel, Dienst ELEM, IIHE, 1050 Brussels, Belgium}

\author{Nick van Eijndhoven}
\affiliation{Vrije Universiteit Brussel, Dienst ELEM, IIHE, 1050 Brussels, Belgium}

\author{Tim Huege}
\affiliation{Institut f\"ur Astroteilchenphysik, Karlsruher Institut f\"ur Technologie, Karlsruhe, Germany}
\affiliation{Vrije Universiteit Brussel, Astrophysical Institute, IIHE, 1050 Brussels, Belgium}
 
\author{Stijn Buitink}
\affiliation{Vrije Universiteit Brussel, Astrophysical Institute, IIHE, 1050 Brussels, Belgium}

\date{\today}

\begin{abstract}
A new generation of neutrino observatories will search for PeV-EeV neutrinos interacting in the ice by detecting radio pulses. Extended air showers propagating into the ice will form an important background and could be a valuable calibration signal. We present results from a Monte-Carlo simulation framework developed to fully simulate radio emission from cosmic-ray particle cascades as observed by in-ice radio detectors in the polar regions. The framework involves a modified version of CoREAS (a module of CORSIKA 7) to simulate in-air radio emission and a GEANT4-based framework for simulating in-ice radio emission from cosmic-ray showers as observed by in-ice antennas. The particles that reach the surface of the polar ice sheet at the end of the CORSIKA~7 simulation are injected into the GEANT4-based shower simulation code that takes the particles and propagates them further into the ice sheet, using an exponential density profile for the ice. The framework takes into account curved ray paths caused by the exponential refractive index profiles of air and ice. We present the framework and discuss some key features of the radio signal and radio shower footprint for in-ice observers.
\end{abstract}

\maketitle


\section{Introduction}

The IceCube neutrino observatory was the first experiment to detect a cosmic high-energy neutrino flux and indicate possible sources, which has been a true milestone in multimessenger astronomy~\cite{aatsen_etal2013, IceCube2018a, IceCube2018b, IceCube2022}. However, IceCube rapidly runs out of events above $\sim$PeV energies, as its current effective volume is too small to address the corresponding very low fluxes. 

Radio signals have an attenuation length 10 times longer than optical light in polar ice, making them a favorable means to monitor very large interaction volumes, and thus enable ultra high energy (UHE) neutrino detection in the $>$~PeV energy range~\cite{ALLISON2012457, Barrella:2010vs, Besson:2007jja, Barwick2005, Avva_2015}. Current in-ice radio neutrino detection experiments with typical peak sensitivities in the higher PeV to EeV energy range consist of two classes. The first class of experiments includes projects like ARA \cite{ARA_paper}, ARIANNA \cite{ARIANNA_paper}, and RNO-G \cite{RNO_paper}. These experiments have deployed radio antennas inside the Antarctic and Greenlandic ice sheets and aim to detect coherent radio emission, known as Askaryan emission, from particle cascades induced by neutrino interactions in ice~\cite{askaryan1962, ZHS1992, Alvarez1997}.

The second class of experiments aims to detect UHE neutrino particle cascades using radar echoes. The in-ice particle cascades are illuminated with radio waves, and the reflected signal is recorded, which could potentially cover the PeV-EeV energy gap. The T-576 experiment at SLAC recently demonstrated that the radar echo method works for particle cascades in High-Density PolyEthylene, a material having very similar properties to polar ice \cite{RET_Obs}. Currently, the Radar Echo Telescope for Cosmic Rays has been deployed at Summit Station, Greenland, as a pathfinder experiment to test the radar echo method in nature~\cite{RET_CR}.

For completeness we mention that alternatives to detecting the radio emission from neutrino interactions with arrays in ice are being explored as well. Examples are balloon-borne radio detectors monitoring large areas of ice such as ANITA~\cite{ANITA} and PUEO~\cite{PUEO}, and radio arrays aiming to detect neutrino-induced air showers like GRAND~\cite{GRAND} and BEACON~\cite{BEACON}.

Radio emission from UHE Cosmic Ray (UHECR) particle cascades acts as an important background for in-ice radio neutrino detectors, as UHECRs have a much higher interaction probability and arrive with a considerably higher flux than UHE neutrinos. Furthermore, they can serve as an important proof of concept and could be an interesting calibration source. It is therefore critical to understand and properly simulate the radio emission from cosmic-ray particle cascades as seen by in-ice detectors. To fully understand the properties of UHECR radio emission, not only the emission during the development of the particle cascade in air has to be taken into account. Especially for detectors deployed at high altitudes, such as at the South Pole or near the summit of the Greenland ice sheet, the emission created during the propagation of the particle cascade through the ice should be considered as well, since at these altitudes the particles in the shower cascade still carry a significant fraction of the energy of the primary particle.

In a previous work, we have presented simulation results of the propagation of UHECR air showers in ice~\cite{DeKockere2022}. We used the CORSIKA Monte Carlo code~\cite{corsika} to simulate air showers, and developed a module based on the GEANT4 simulation toolkit~\cite{Geant4} to propagate the air shower particle footprints through a high-altitude ice layer. We showed that the particle showers contain a very energy-dense core, forming the main component determining the development of the in-ice part of the cascades. We found that the lateral charge distribution of the in-ice cascades has a comparable typical width to that of neutrino-induced particle cascades, indicating that the radio emission from both types of events shares many similarities. We implemented the endpoint formalism~\cite{corsika_endpoint} within the GEANT4 module to get a first estimation of the Askaryan emission from the in-ice particle cascades, ignoring detailed radio signal propagation in the ice. We concluded that the energy-dense core containing the particles of the cascade up to a radius of $1$~m dominates the radio emission, leading to similar signals compared to neutrino-induced particle cascades.

Following up on the previous work summarized above we now present FAERIE, the Framework for the simulation of Air shower Emission of Radio for in-Ice Experiments. It is the first complete Monte-Carlo cosmic-ray radio emission simulation framework for in-ice detectors. It includes both the propagation of the particle cascade in air and in ice, and applies the endpoint formalism to calculate the radio emission from both components using ray tracing to account for radio propagation through nonuniform media.

\section{Framework overview}

To fully simulate the radio emission of cosmic-ray particle cascades for in-ice radio antennas, we need to consider: a) the emission created during the development of the in-air particle shower and how this propagates into ice, and b) the emission created during the development of the in-ice particle cascade, initiated when the particle shower itself propagates into ice. Earlier studies investigating the propagation of cosmic-ray particle cascades into ice and other media can be found in~\cite{ZHS1992, Razzaque2002, Razzaque2002_add, Bevan_2007, Seckel2008, Tueros2010, Alvarez_2012, Javaid2012, SAFTOIU_2013, deVries2016, DeKockere2021, DeKockere2022, DeKockere2023}. Polar ice can reach up to altitudes of $\sim 3$~km, which corresponds to a vertical atmospheric depth of $\sim 730$~g/cm$^2$ at the South Pole, where UHECR air showers still have a very energy-dense core. As such, the in-ice radio emission can lead to a significant contribution to the total radio emission of the shower as observed by in-ice antennas.

FAERIE uses the air shower simulation program CORSIKA 7.7500 to simulate the in-air particle cascade, which calculates the radio emission using the CoREAS code~\cite{corsika, MainCoreas}. CoREAS applies the endpoint formalism to simulate the electric field at a given antenna position in air. This is done by calculating the emission of every single charge in the particle cascade during the CORSIKA simulation, and using straight-line propagation of rays \cite{corsika_endpoint,SLAC_T510}. For FAERIE, we modified the code so it can handle antenna positions in ice using full ray tracing. The ray tracing has been discussed in detail in ~\cite{Latif2023, Latif_ICRC23}.

For the simulation of the in-ice particle cascade, FAERIE uses a code based on the GEANT4 simulation toolkit \cite{Geant4}. GEANT4 allows for full Monte-Carlo simulation of particle creation and propagation through any given medium. Using GEANT4 we propagate the core of the in-air particle shower, i.e. all particles within $1$~m of the shower axis, into an ice volume consisting of multiple horizontal 1-cm-thick layers of pure ice. Figure~\ref{fig:Ecum_comb} shows the energy contained within a given radius from the shower core for two simulated proton-induced air showers at an altitude of $2.835$~km, which corresponds to a vertical atmospheric depth of $729$~g/cm$^3$. Both showers were simulated using a zenith angle $\theta = 0$ but differ by one decade in primary energy. For the shower with a primary energy of $E_p = 10^{17}$~eV we see that at the given altitude the full air shower contains roughly $50$\% of the primary energy. Within a radius of $1$~m from the shower core we find about $20$\% of the primary energy, which is about half of the energy of the full particle shower at this altitude. A similar picture holds for the shower with a primary energy of $E_p = 10^{18}$~eV, where the full shower contains close to $60$\% of the primary energy and again half of that energy is contained within $1$~m of the shower axis. As illustrated by Figure~\ref{fig:E_dens_both} in Appendix~\ref{App:shower_profiles}, propagating the showers through ice at the given altitude leads to high energy densities in the ice close to the shower core, dropping rapidly as the distance to the shower axis increases. Similar results were shown in Figures~5~and~6~in~\cite{DeKockere2022}.  Furthermore, Figure~16 in~\cite{DeKockere2022} shows that we do not expect any significant changes in the total fluence of the in-ice radio emission around the Cherenkov cone, when the radius of the particle footprint being propagated into the ice is increased beyond $\sim 10$~cm. Important to note is that only for the in-ice cascade the simulation is limited to a radius of $1$~m from the shower axis. For the simulation of the in-air particle cascade and its corresponding radio emission, the whole air shower is accounted for.

Each layer of the ice has a constant density determined by the depth of the layer, following a typical polar ice density profile given by
\begin{equation}\label{Eq:ice_dens}
    \rho(z) = \rho_{\text{ice}} - (\rho_{\text{ice}} - \rho_{\text{surface}}) \exp\left(-\frac{1.9}{t_{\text{firn}}} |z|\right),
\end{equation}
with $\rho_{\text{ice}} = 917$~kg~m$^{-3}$, $\rho_{\text{surface}} = 359$~kg~m$^{-3}$ and $t_{\text{firn}} = 100$ m \cite{kravchenko_besson_meyers_2004}. The ice density profile can be freely adjusted by the user. The radio emission is simulated using the endpoint formalism, using the implementation originally developed for the T-510 experiment at SLAC~\cite{Bechtol_2022}, with the addition of full ray tracing to account for the changing index of refraction in the ice. The index of refraction profile is not directly related to the layered density profile described by Equation~\ref{Eq:ice_dens}, but instead follows a continuous exponential profile given below. Note that this means that the layered density profile of the ice volume only influences the development of the particle cascade, and does not influence the radio propagation. The features of the density profile at depths beyond $\sim 20$~m are therefore not relevant. The GEANT4-based particle propagation program has been discussed in detail in \cite{DeKockere2022, DeKockere2023}.

As shown in~\cite{corsika_endpoint}, the endpoint formalism naturally includes the emission of transition radiation when the calculations of the emission in the first medium and the emission in the second medium are performed separately, which is how it is implemented in FAERIE.

\section{Ray tracing}

\begin{figure}
\includegraphics[width=\linewidth]{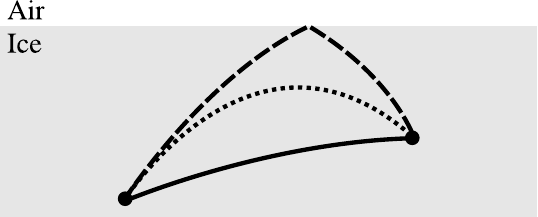}
\caption{\label{fig:diff_rays} A sketch of the three different types of rays: a direct ray (solid line), an indirect refracted ray (dotted line) and an indirect reflected ray (dashed line). In reality, only one of the two indirect rays will be a solution of the ray tracing.}
\end{figure}

To account for the changing index of refraction in both air and ice, as well as the transition of the radiation from air to ice, FAERIE uses full ray tracing for the simulation of the radio emission. Ray tracing is the procedure of tracing the trajectory of radio waves traveling through a given medium, and/or passing the boundary of two different media. Ray tracing in the simulation is performed analytically in a flat Earth approximation using an exponential refractive index profile for both air and ice. Furthermore we assume a cylindrical symmetry for both media, i.e. the refractive index profile only depends on the depth in the air or ice. The air density profile and corresponding refractive index profile as well as the refractive index profile for the ice can be freely adjusted by the user.

For the results presented in this work, the density of the atmosphere is described by a five-layer model for the South Pole fitted to a database used by the National Center for Environmental Predictions Global Forecast System in weather forecasting. Based on this fit, a five-layer model of the refractive index is created. Both the five-layer density model as well as the corresponding five-layer refractive index model are then used for the CORSIKA/CoREAS simulation~\cite{Latif2023}. The details of the models are given in Appendix~\ref{App:atmosphere}. In the following, the refractive index model for ice is set to
\begin{equation}\label{eq:index_profile}
    n(z) = 1.78 - 0.43 \exp\left(-(0.0132 \text{ m}^{-1}) |z|\right),
\end{equation}
corresponding to the model used by the Askaryan Radio Array at the South Pole outlined in~\cite{Kelley2018}. More detailed information on the analytical ray tracer can be found in \cite{Latif_thesis_2020, Latif_IceRayTracing_2020}.

Air-to-ice ray tracing describes ray bending in air, followed by refraction at the air-ice boundary and further bending in the ice. The ray-bending effect is less pronounced in air than in ice since the air refractive index value remains close to unity throughout. The refractive index profile is however still relevant, as the refractivity $R = n - 1$ starts around $R = 0$ at the top of the atmosphere and increases to approximately $R = 3 \times 10^{-4}$ at sea level, which can lead to a so-called refractive displacement for inclined air showers~\cite{Schluter:2020tdz}. The polar ice refractive index changes from $n = 1.35$ at the surface to $n = 1.78$ within the first $100-200~$m. Therefore, significant refraction occurs near the ice surface due to a sharp change in the refractive index values. For ice-to-ice ray propagation, in general two solutions are found for each emitter-receiver pair. The first solution we will call the direct ray, which out of the two solutions has the shortest ray path between emitter and receiver. The second solution we will call the indirect ray, which can either be a refracted ray with a ray path longer than that of the direct ray, or a ray reflecting on the ice-air boundary. All three types of rays are shown in Figure~\ref{fig:diff_rays}.

In order to obtain ray parameters like the ray propagation time and path length from the analytic ray tracing expressions, one must provide the initial launch angle of the ray. At the correct launch angle, the ray will hit the target point in ice. Therefore, ray tracing involves a minimization procedure. This procedure will be slightly different from when we are tracing rays from a point in air to a point in ice as compared to when the rays are being traced between two in-ice points. 

\subsection{Interpolation of ray tracing parameters}

A typical execution of the analytic ray tracing functions involving air and ice takes around $0.05-0.1$ ms. This is relatively fast but not fast enough for simulating a complete cosmic ray shower. A typical cosmic-ray shower will consist of $\mathcal{O}(10^9)$ particles at EeV energies. In this case, propagating rays from each shower particle at each step to each antenna could take weeks or months, depending on the shower energy and the number of in-ice antennas. Therefore, it is not feasible to use the analytic ray tracing functions directly, and we have to move toward interpolation tables. Interpolation of ray tracing parameters from premade tables makes the ray tracing process significantly faster.

For each receiver point of interest, an interpolation table is generated by using a 2D grid representing the particle shower, either in air or in ice. The grid covers the expected cascade dimensions, using horizontal distance with respect to the receiver point on the first axis, and depth of cascade on the second axis. The points in the grid are considered the emitter points, and ray tracing is performed between each emitter point and the given receiver point. For each grid point the resulting ray tracing parameters are stored in a table, leading to one single table for each receiver. 

To reduce memory usage during the simulation, antennas at the same depth in the ice share a single interpolation table for the in-air ray tracing. Since the index of refraction profile only depends on depth, the ray tracing applied is cylindrically symmetric. A grid covering a sufficiently large interval of horizontal distances can therefore be used by multiple antennas located at the same depth in the ice. For the in-air ray tracing the grid depth is varied from the ice layer altitude, generally around 3~km, up to $100~$km. The horizontal distance range for the grid is set by varying the ray launch angle $\theta$ from $89.9^{\circ}$ (almost horizontal) to $0^{\circ}$ (vertically down). Small variations in $\theta$ can cause almost exponential changes in the horizontal distance. 

\begin{figure}
\includegraphics[width=\linewidth]{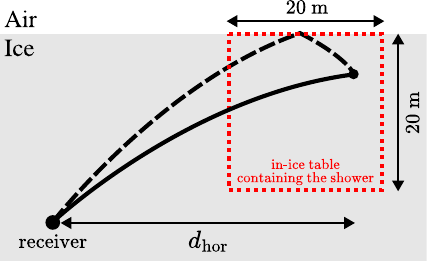}
\caption{\label{fig:interpol_table} A visualization of an interpolation table used for the in-ice ray tracing, for a given receiver position. Interpolation tables for the in-air ray tracing follow a similar structure, but instead cover a much larger area in air. A single in-air table is used for multiple receivers at the same depth in the ice.}
\end{figure}

The in-ice ray tracing interpolation tables span intervals of 20~m on both axes, which covers the typical dimension of in-ice cascade. Since in ice the refractive index changes much more than it does in air, the in-ice grids are denser than the in-air grids. For in-ice ray tracing, one interpolation table is made for every single antenna, irrespective of the antenna depths in the ice. A visualization of an interpolation table used for the in-ice ray tracing is given in Figure~\ref{fig:interpol_table}.

Once generated, the tables are used to supply the CoREAS and GEANT-based modules with the ray parameters for any given emitter point they encounter by using linear interpolation between the grid values. Each interpolation for a given parameter takes around $250~$ns, making the ray tracing process significantly faster and, as such, particle shower radio emission calculations in gradient media feasible. The error for the interpolated results is around $\mathcal{O}(10^{-7}-10^{-8})$ \cite{Latif_ICRC23,Latif2023}.

Note that the method of using interpolation tables for ray tracing is independent of the type of ray tracer being used. Since they are created before the actual simulation starts, also slower and more involved ray tracers could be considered for setting up the interpolation tables.

\subsection{Modification of the endpoint formalism}

\begin{figure*}
\includegraphics[trim={0 0 0 0},clip,width=0.45\linewidth]{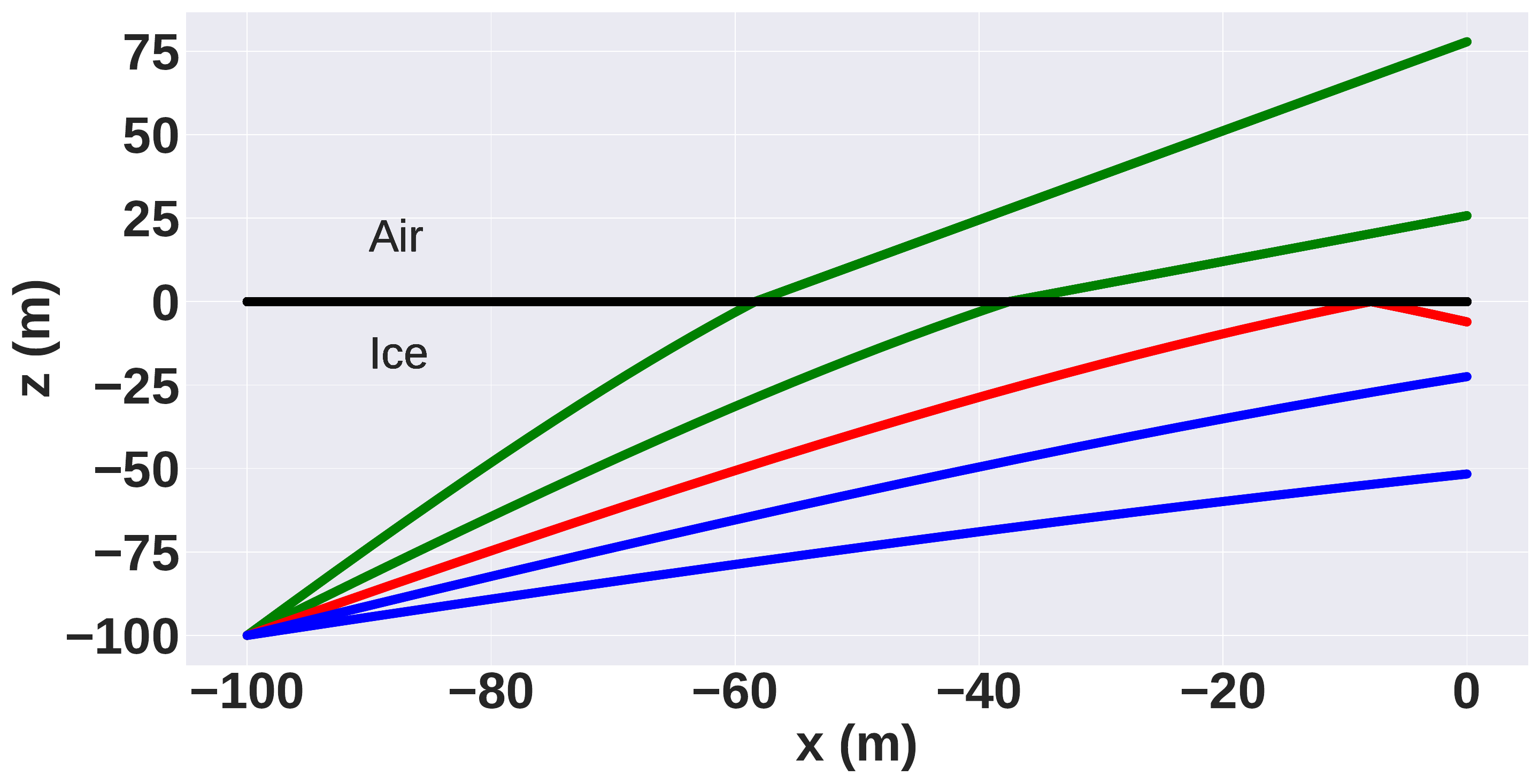}
\includegraphics[trim={0 0 0 0},clip,width=0.45\linewidth]{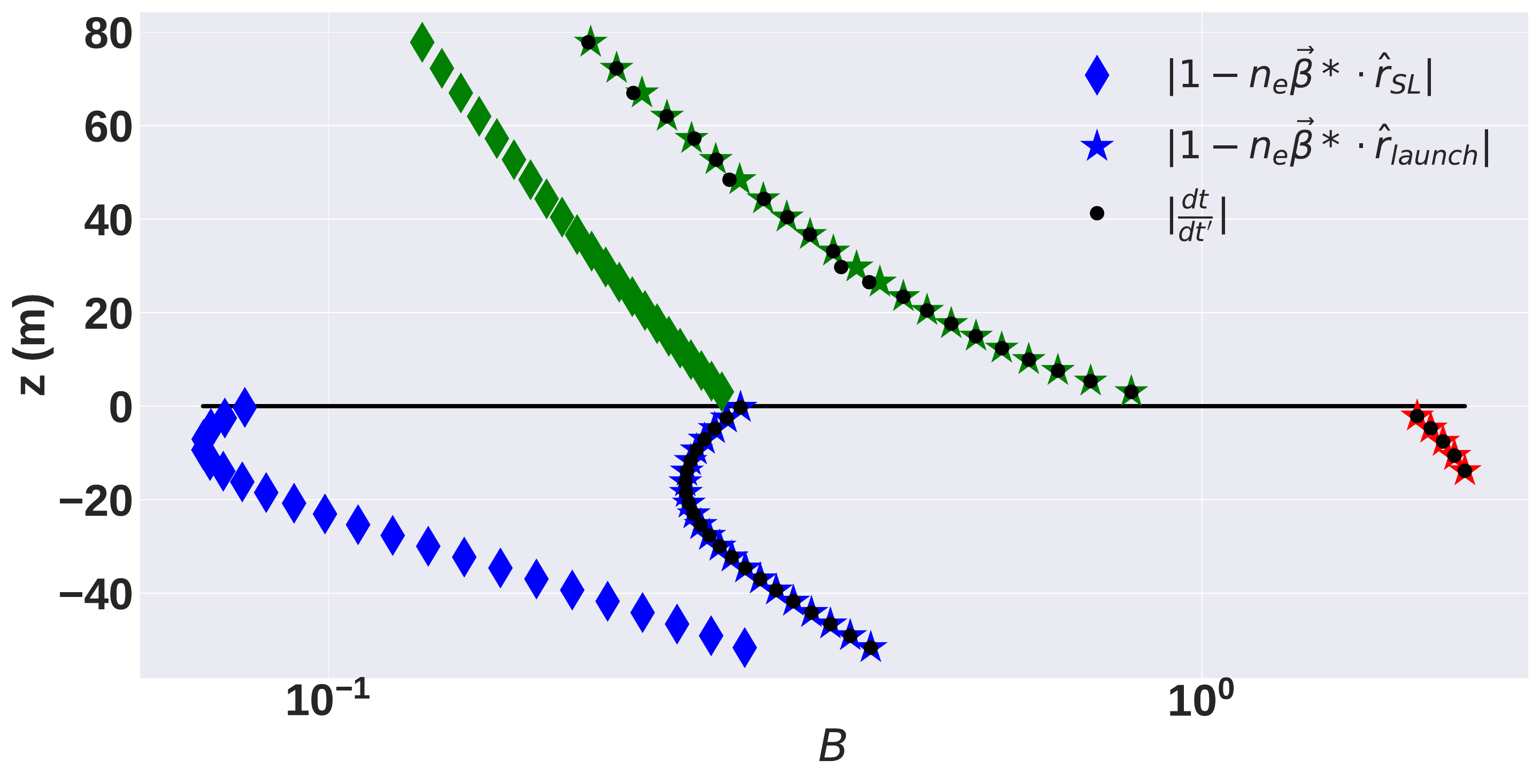} 
\caption{\label{fig:RayPathbf}} Left: Visualization of several rays traced from the receiver to a line of emitters, including an air-ice boundary at $z = 0$. Direct rays are shown in blue. Reflected rays are only considered when total internal reflection occurs and are shown in red. Transmitted rays are shown in green. Right: the boost factor estimations along with the numerically computed value of $\frac{dt}{dt'}$ as a function of depth. The diamonds correspond to the calculations using $\hat{r}$ pointing directly from
emission point to receiver point. The stars correspond to the values of the boost factor when evaluating $n$ at the emission point and using the launch direction for $\hat{r}$. The dots correspond to the numerically calculated values of the boost factor. The different regions are associated with direct (blue), reflected (red) and transmitted rays (green).
\end{figure*}

The endpoint formalism calculates the electric field observed at a position $\vec{x}$ emitted by a charged particle taking a propagation step (track segment) in the simulation by using the formula

\begin{eqnarray}\label{eq:End-point}
\vec{E}_{\pm}(\vec{x},t) = \pm \frac{1}{\Delta t} \frac{q}{c}\left( \frac{\hat{r} \times [\hat{r} \times \vec{\beta}^*]}{|1-n\vec{\beta}^* \cdot \hat{r}|R} \right),
\end{eqnarray}

where the plus sign is applied using the start point of the step and the minus sign using the end point of the step~\cite{corsika_endpoint}. The combination of both then gives the net contribution of the step. Here, $\Delta t$ is the sampling time interval of the observer, $q$ the charge of the particle, $\hat{r}$ the direction from the start/end point toward the antenna, $\vec{\beta}^{*}$ the velocity of the charge during the step, $R$ the distance between the start/end point and the antenna, and $n$ the index of refraction of the medium at the respective emission point. The variable $t$ refers to the arrival time of the emission at the observer, and relates to the emission time $t'$ via $t = t' + nR/c$. The emission time $t'$ is often referred to as the retarded time.

The so-called boost factor $B = 1-n\vec{\beta}^* \cdot \hat{r}$ found in Equation~\ref{eq:End-point} arises from classical theory and can more generally be described as $B = \frac{dt}{dt'}$. The implementation as used in the endpoint formalism considers signals traveling along straight lines, i.e. $\hat{r}$ points directly from emission point to receiver point. This approach is however not valid when working in media with a nonconstant index of refraction where formalisms such as ray tracing need to be used to correctly describe signal propagation.

Using a ray tracer calculating the derivative $\frac{dt}{dt'}$ numerically, we found that the relation $B = 1-n\vec{\beta}^* \cdot \hat{r}$ still holds provided that $n$ is evaluated at the emission point and $\hat{r}$ is interpreted as the launching direction of the ray~\cite{VandenBroeck2023}. Figure~\ref{fig:RayPathbf} shows a visualization of the ray tracing together with the comparison of different estimators of the boost factor $B$ to the numerically computed value of $\frac{dt}{dt'}$. Rays were traced from a receiver in the ice to a line of emitting points, including an air-ice boundary at $z=0$. Both the in-air and the in-ice medium have an associated exponential index of refraction profile. Evaluating the boost factor using a straight line calculation clearly diverges from the numerically calculated value, while evaluating $n$ at the emission point and using the launch direction for $\hat{r}$ agrees well for all direct, reflected and refracted rays.

The derivation in~\cite{ZHS1992} shows that the variable $R$ in Equation~\ref{eq:End-point} refers to the geometrical distance between the emitter point and the receiver point. The variable $R$ is therefore now interpreted as the geometrical path length of the ray connecting emitter and receiver, which implies the relation $t = t' + nR/c$ still holds.

\begin{figure}
\includegraphics[width=0.7\linewidth]{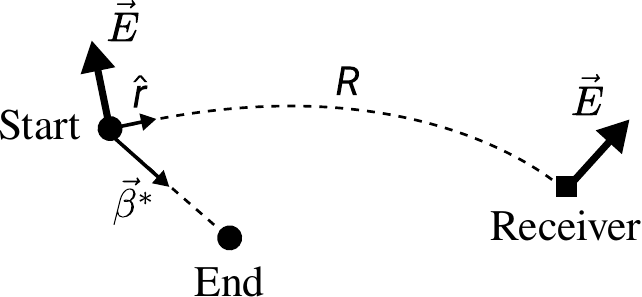}
\caption{\label{fig:End-point} A visualization of the variables used in the endpoint formula including ray tracing (Equation~\ref{eq:End-point}), in the specific case where the start point, end point and receiver coincide in the same plane. The circles represent the start and end point of the emitter step, the square represents the receiver. All vectors shown in the illustration lie in the same plane.}
\end{figure}

Finally, a rotation of the electric field is applied in the plane of the ray. This rotates the electric field such that its component in the plane of the ray is perpendicular to the receiver direction, instead of the launching direction. A visualization of the endpoint formalism including ray tracing is shown in Figure~\ref{fig:End-point}.

\subsection{Fresnel coefficients}

In order to correctly calculate the final electric field amplitude as observed by the in-ice antenna, it is important to calculate Fresnel coefficients for the electric field ray paths. In case of air-to-ice rays moving through the air-ice boundary, we need to apply the transmission coefficients. In case of ice-to-ice rays reflecting on the ice-air boundary, we need to apply the reflection coefficients.

\begin{figure}
\includegraphics[width=0.8\linewidth]{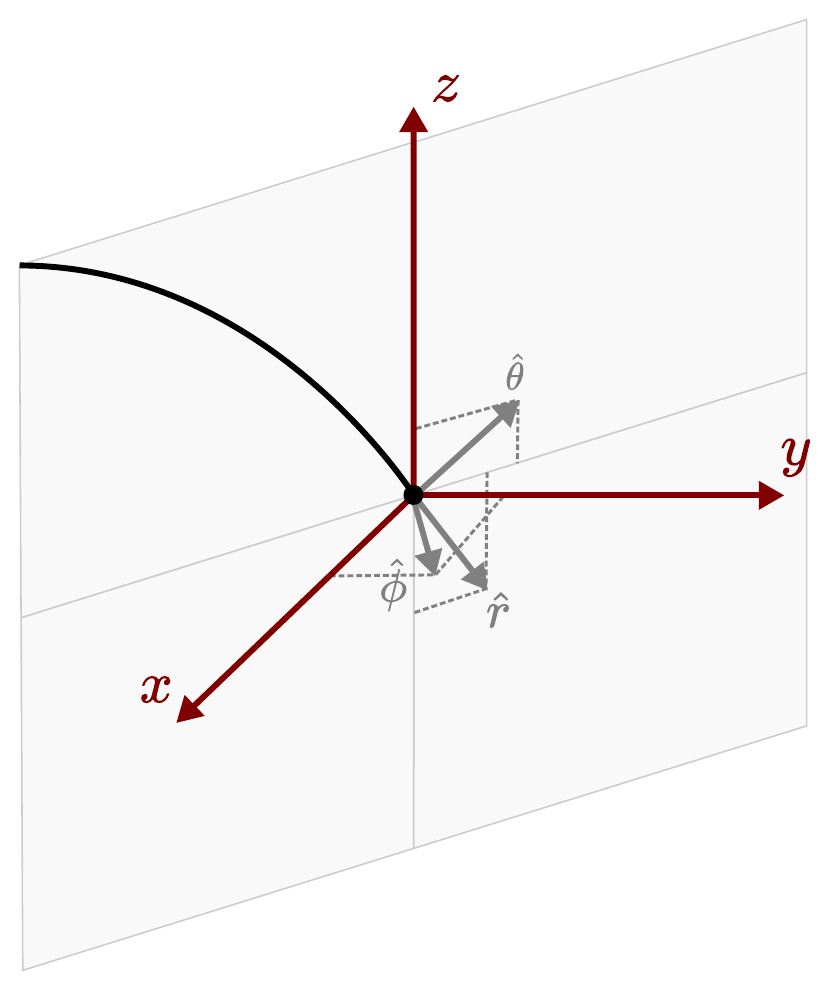}
\caption{\label{fig:Fresnel_coords} A visualization of the relation between a global Cartesian coordinate system and a local spherical coordinate system, defined by the incoming direction of the ray.}
\end{figure}

The cosmic-ray simulation framework utilizes a global Cartesian coordinate system in which the positions and momenta of the particles; the positions of the receivers; and the electric field components $E_x$, $E_y$, and $E_z$ are defined. To apply the Fresnel coefficients we have to construct the orthogonal vectors $\hat{r}$, $\hat{\theta}$, and $\hat{\phi}$ representing a local spherical coordinate system defined by the incoming direction of the ray, within the global Cartesian coordinate system of the simulation framework. This is shown in Figure~\ref{fig:Fresnel_coords}. The Fresnel coefficients are commonly expressed as scaling factors for the S and P electric field components. The P component is the component in the plane of incidence of the ray, i.e. the plane of the ray, while the S component is the component perpendicular to that plane. Since the electric field is perpendicular to the propagation direction and therefore to the unit vector $\hat{r}$, we know that $E_{\text{S}} = E_{\phi}$ and $E_{\text{P}} = E_{\theta}$.

From ray tracing we find the propagation direction of the ray and, therefore, can construct the vector $\hat{r}=(r_x,r_y,r_z)$ in the Cartesian basis. Since we know that $\hat{\phi} \cdot \hat{r} = 0$ and $\hat{\phi}$ is confined to the $xy$ plane, we find
\begin{equation}
    \hat{\phi}=\frac{1}{\sqrt{r_x^2+r_y^2}}\left(-r_y,r_x,0\right).
\end{equation}
The unit vector $\hat{\theta}$ can then be found by $\hat{\theta} = \hat{\phi} \times \hat{r}$, which gives
\begin{equation}
    \hat{\theta} = \frac{1}{\sqrt{r_x^2 + r_y^2}} \left( r_x r_z, r_y r_z, -(r_x^2 + r_y^2) \right)
\end{equation}

The electric field $\vec{E}_C$ corrected for the transmission coefficients can therefore be derived from the uncorrected electric field $\vec{E}$ using
\begin{equation}\label{eq:correction}
    \vec{E}_C = (\hat{r} \cdot \vec{E}) \hat{r} + t_{\text{S}} (\hat{\phi} \cdot \vec{E}) \hat{\phi} + t_{\text{P}} (\hat{\theta} \cdot \vec{E}) \hat{\theta}.
\end{equation}
The correction for the reflection coefficients follows the same formula, where the transmission coefficients $t_{\text{S}}$ and $t_{\text{P}}$ are replaced by respectively the reflection coefficients $r_{\text{S}}$ and $r_{\text{P}}$.

As the Fresnel coefficients are scaling factors for the S and P components of the electric field, the correction given by Equation~\ref{eq:correction} can be applied directly to the electric field at the receiver, using $\hat{r}$, $\hat{\phi}$, and $\hat{\theta}$ in the local spherical coordinate system at the receiver. There is no need to calculate the electric field and the unit vectors of the corresponding local spherical coordinate system at the point where the ray meets the air-ice boundary. The effect of propagation from the air-ice boundary to the receiver on the electric field is, following Equation~\ref{eq:End-point}, limited to an overall decrease in magnitude by a given factor due to the longer geometrical path length $R$, followed by a rotation in the plane of the ray. The only relevant variable at the air-ice boundary is the incidence angle $\theta_i$, i.e., the angle between the refracting or reflecting ray and the normal on the interface, which follows directly from ray tracing.

Both the permeabilities in air and ice can be approximated by the permeability of free space $\mu_0$, and the coefficients therefore follow the relations given in~\cite{Hecht2002}. Using Snell's law they can be expressed as
\begin{subequations}
\label{eq:Fres_Eqns}
\begin{equation}
r_\text{S}=\frac{n_1\cos(\theta_i)-n_2\sqrt{1-\left(\frac{n_1}{n_2}\sin(\theta_i)\right)^2}}{n_1\cos(\theta_i)+n_2\sqrt{1-\left(\frac{n_1}{n_2}\sin(\theta_i)\right)^2}}\label{subeq:S_reflactance}
\end{equation}
\begin{equation}
    t_\text{S}=1+r_\text{S} \label{subeq:S_transmittance}
\end{equation}
\begin{equation}
r_\text{P}=-\frac{n_1\sqrt{1-\left(\frac{n_1}{n_2}\sin(\theta_i)\right)^2}-n_2\cos(\theta_i)}{n_1\sqrt{1-\left(\frac{n_1}{n_2}\sin(\theta_i)\right)^2}+n_2\cos(\theta_i)}\label{subeq:P_reflactance}
\end{equation}
\begin{equation}
    t_\text{P}=(1+r_\text{P})\frac{n_1}{n_2} \label{subeq:P_transmittance}
\end{equation}
\end{subequations}
The variables $n_1$ and $n_2$ refer to the refractive index values of the first and the second medium at the boundary respectively. The ordering of the two media (i.e., air and ice) can differ depending on whether a ray is being transmitted through the air-ice boundary or reflected on the ice-air boundary. It should be noted that for $\theta_i$ greater than the angle of total reflection a nontrivial phase shift occurs, which is not included in this work.

\subsection{Focusing factor}

As a consequence of ray tracing, rays will converge or diverge when compared to the straight-line approximation. Since during the simulations the emitter and receiver are represented by points and therefore do not capture any convergence or divergence effects during the ray tracing, this needs to be taken into account explicitly. We follow the same procedure as described in~\cite{NuRadioMC}, which introduces a focusing factor correcting the electric field amplitude. This factor is given by 
\begin{equation} \label{eq:Focusing_Factor}
F=\sqrt{\frac{n_{Tx}}{n_{Rx}}\frac{R}{\sin{\theta_{Rec}}}\frac{\Delta \theta_{L}}{\Delta z}},
\end{equation}
and is calculated for the ice-to-ice direct and indirect ray tracing. Here $n_{Tx}$ and $n_{Rx}$ are the refractive index values at the transmitter and receiver points respectively, $R$ is the geometrical path length of the ray between the two points, $\theta_{Rec}$ is the angle of reception at the receiver point, and $\Delta \theta_{L}$ is the difference in ray launch angles ($\theta_{L}$) for when the receiver is moved away from its ``original'' depth by an amount given by $\Delta z$. In our work, $\Delta z$ was fixed to $1~$cm. Following the approach in~\cite{NuRadioMC}, we limit the factor to $ 0.5 \leq F \leq 2.0$.

In principle a focusing factor should also be applied to rays moving from air into ice, and a method similar to the one described in~\cite{Lehtinen_2004} could be used. However, this effect is negligible when the angle of incidence of the ray to the surface normal is small. Since the Cherenkov angle in air is of the order of $1^{\circ}$, we do not include it in the simulations. We expect it to be important only for inclined air showers, at which point also the curvature of the Earth needs to be taken into account.

\section{Simulation Results}

\begin{figure*}
    \centering
    \includegraphics[width=\textwidth]{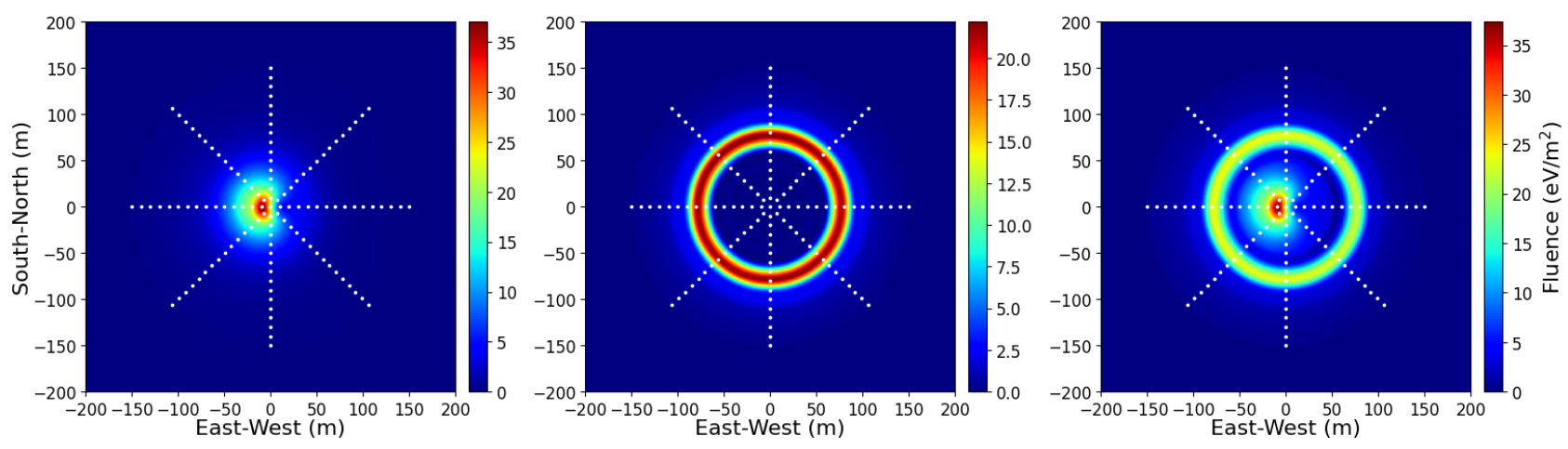}
    \caption{The fluence footprint of the simulated cosmic-ray shower at a depth of 100~m for the in-air emission (left), in-ice emission (middle), and the combined emission (right), using a primary energy of $E_p = 10^{17}$~eV and zenith angle $\theta = 0$. The simulation was performed using 121 antennas in a star-shaped grid with 8 arms and an antenna spacing of 10~m, indicated by the white dots. For the interpolation of the star grid we used the code described in~\cite{Corstanje_Code}.}
    \label{fig:fp_combined}
\end{figure*}

\begin{figure*}
    \centering
    \includegraphics[width=0.9\textwidth]{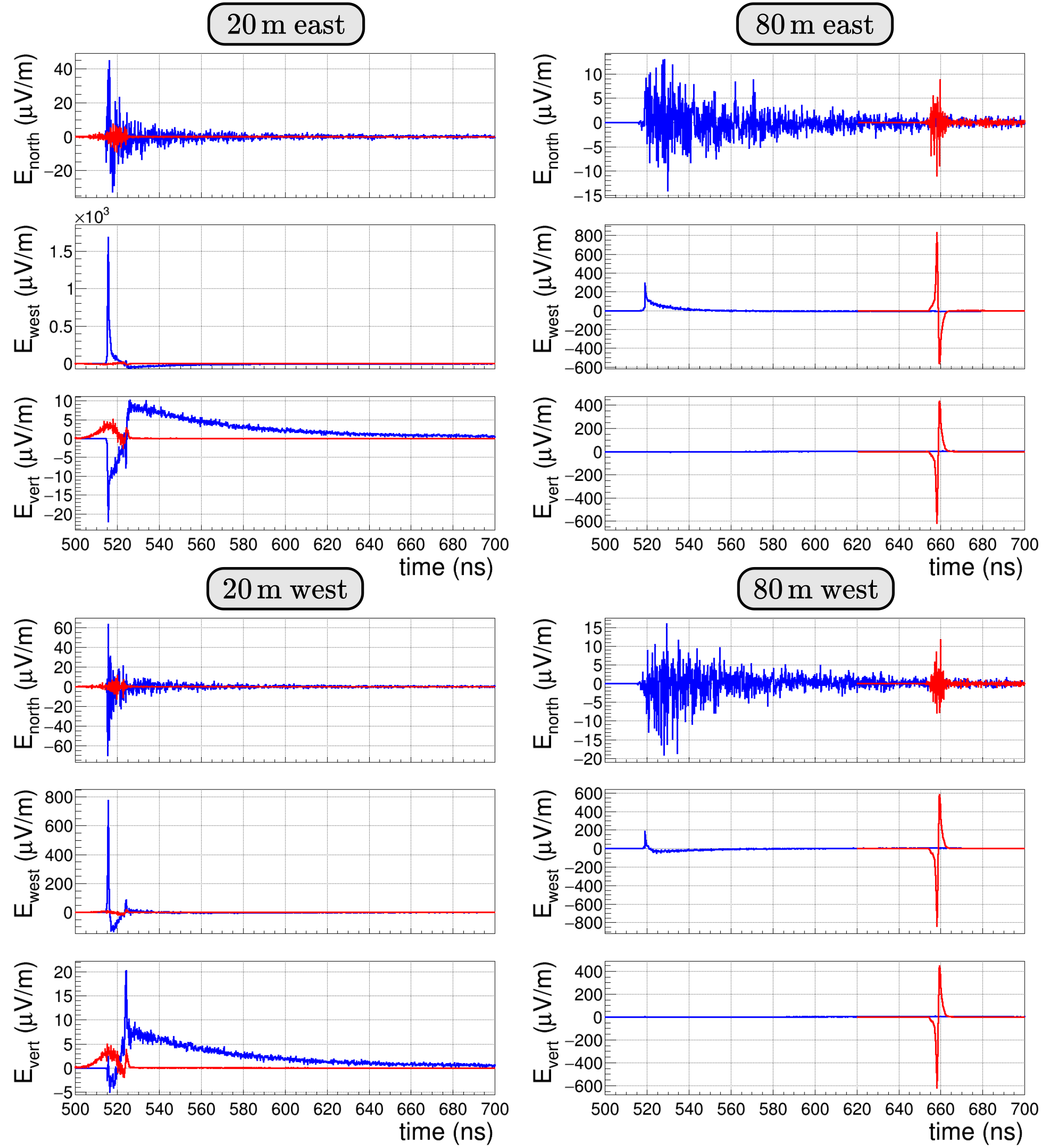}
    \caption{The electric field components as a function of time for 4 different antennas positioned along the west axis at a depth of 100~m, using a primary energy of $E_p = 10^{17}$~eV and zenith angle $\theta = 0$. The coordinate of each antenna is indicated at the top of the plots. The blue line indicates the in-air emission, and the red line indicates the in-ice emission. Note that the range on the y axis is different for every plot.}
    \label{fig:traces_west_axis_comb}
\end{figure*}

\begin{figure*}
    \centering
    \includegraphics[width=0.9\textwidth]{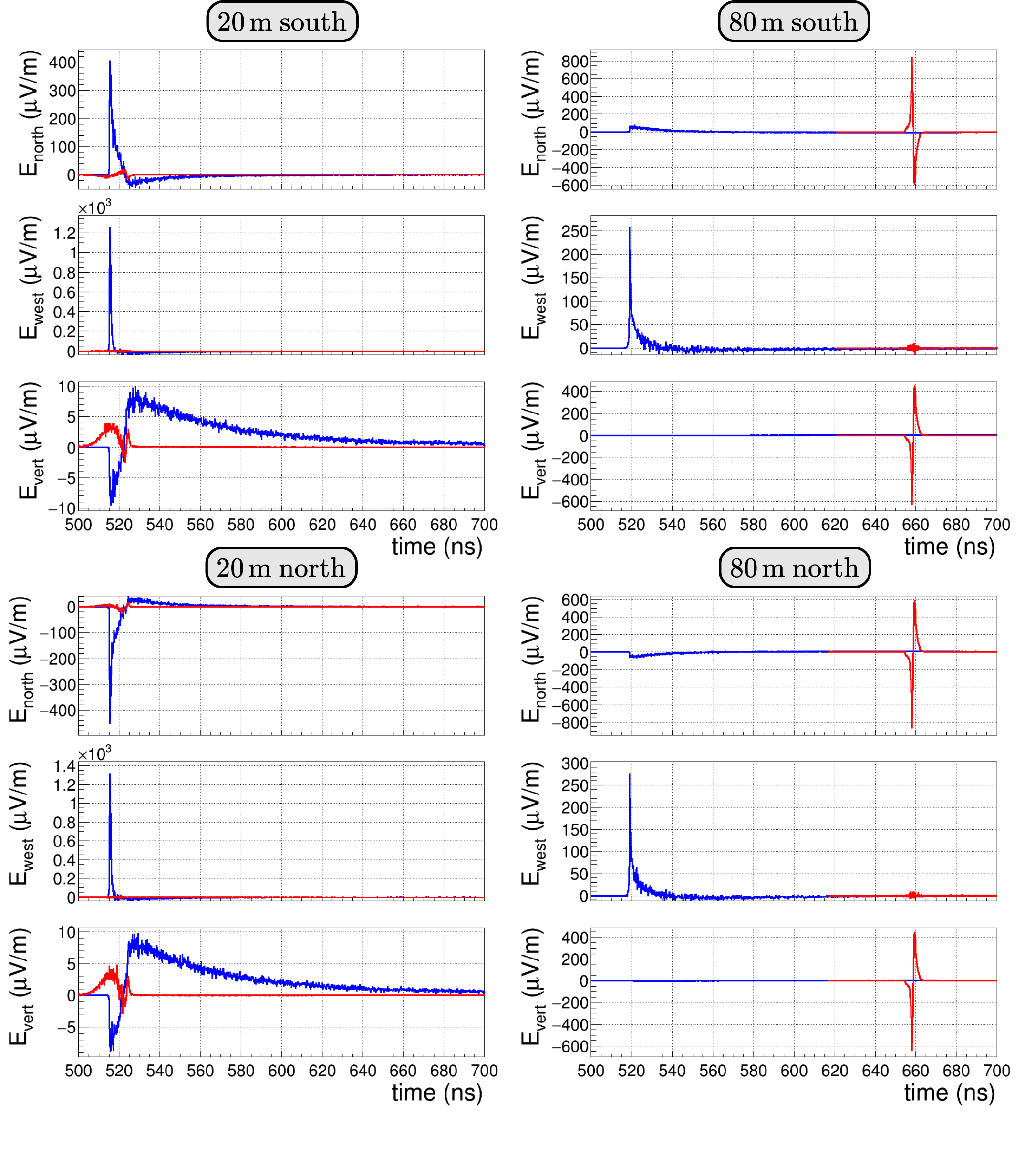}
    \caption{The electric field components as a function of time for 4 different antennas positioned along the north axis at a depth of 100~m, using a primary energy of $E_p = 10^{17}$~eV and zenith angle $\theta = 0$. The coordinate of each antenna is indicated at the top of the plots. The blue line indicates the in-air emission, and the red line indicates the in-ice emission. Note that the range on the y axis is different for every plot.}
    \label{fig:traces_north_axis_comb}
\end{figure*}

In this section we discuss simulation results from FAERIE. We simulated a cosmic-ray cascade initiated by a $E_p = 10^{17}$~eV proton entering the atmosphere at a zenith angle $\theta = 0^{\circ}$ and azimuth angle $\phi = 0^{\circ}$. We set the geomagnetic field to $\vec{B} = (16.7525\text{ }\mu\text{T}, -52.0874\text{ }\mu\text{T})$, which are the values at a latitude of $89.9588^{\circ}$ south following the IGRF12 model~\cite{magn_field}. The first component indicates the horizontal component of the magnetic field and defines the geomagnetic north. The second component indicates the vertical component of the magnetic field, and is upward for negative values. Thinning was applied during the CORSIKA air shower simulation on electromagnetic particles falling below $10^{-6} E_p$, using a weight limit of $w = 100$~\cite{Heck1998, KOBAL2001259}. For hadrons (except $\pi^0$'s) and muons the kinetic energy cutoff was set to $0.3$~GeV. For electrons and photons (including $\pi^0$'s) the kinetic energy cutoff was set to $0.401$~MeV. We use the QGSJETII-04 high energy hadronic interaction model~\cite{QGSJET} and the UrQMD low energy hadronic interaction model~\cite{UrQMD_1, UrQMD_2}. The altitude of the ice layer was fixed to that of the South Pole, which is $2.835$~km and corresponds to a vertical atmospheric depth of $729$~g/cm$^2$. For the GEANT4 simulation we used the default production cut-off length for gammas, electrons, positrons and protons, which is $1$~mm. Only charged particles with a kinetic energy above $0.1$~MeV were taken into account for the calculation of the radio emission. The longitudinal particle and energy distributions for the in-air particle cascade are shown in Appendix~\ref{App:shower_profiles} (Figure~\ref{fig:part_dist_1e8}), together with the energy deposited in the ice by the in-ice particle cascade (left-hand side of Figure~\ref{fig:E_dens_both}). More information about the CPU time and job run time of the simulation can be found in Appendix~\ref{App:run_time}.

Figure~\ref{fig:fp_combined} shows the fluence footprint of the simulated cosmic-ray cascade at a depth of 100~m below the air-ice boundary, for the in-air emission (left), in-ice emission (middle) and combined emission (right), where fluence is defined as
\begin{equation}\label{Eq:fluence}
    \mathcal{F} = \epsilon_0 c_0 \int E^2(t) \mathrm{d}t,
\end{equation}
with $\epsilon_0$ the vacuum permittivity and $c_0$ the speed of light in vacuum.

The simulation was performed using 121 antennas in a star-shaped grid with 8 arms and an antenna spacing of 10~m, including a single antenna at (0,0). For the interpolation of the star grid we used the code described in~\cite{Corstanje_Code}. We follow the CORSIKA coordinate system, consisting of the geomagnetic north, west and vertical directions. The center of the coordinate system corresponds to the point of impact of the shower core on the ice surface. In this case, west is the polarization direction of the geomagnetic emission ($\vec{v}\times\vec{B}$).

The in-air emission in Figure~\ref{fig:fp_combined} (left) clearly shows the typical bean shape which arises due to the interference between the geomagnetic emission and the Askaryan emission, as described in~\cite{de_Vries_2010}. The in-ice emission is dominated by Askaryan radiation, and therefore a clear Cherenkov-like ring can be seen. The combination of both gives a unique footprint, showing the beanlike shape from the in-air emission close to the shower axis, enclosed by the Cherenkov-like ring from the in-ice emission.

Figure~\ref{fig:traces_west_axis_comb} and~\ref{fig:traces_north_axis_comb} show the electric field components as a function of time for 4 different antennas positioned along the west axis and the north axis respectively, at a depth of 100~m. The blue lines indicate the in-air emission, while the red lines indicate the in-ice emission. The coordinates of the antennas on their given axis are shown at the top of each plot. As shown in~\cite{NuRadioMC}, the typical peak amplitude around the Cherenkov cone of neutrino-induced Askaryan signals from electromagnetic showers in ice with an initial energy of $10^{16}$~eV is of the order of $\sim 1000$~$\mu$V/m, which is similar to the peak amplitudes shown in both figures. The signal of a cosmic ray cascade with a primary energy $E_p$ and zenith angle $\theta = 0$ thus roughly translates to a signal from a neutrino with an energy of $0.1 \times E_p$, which indeed corresponds to the typical energy within the cosmic-ray air shower core at these altitudes~\cite{DeKockere2022}.

As we move further outward, the difference in time of arrival between the in-air and in-ice emission increases. This increase in delay can be expected. For antennas further away from the point of impact of the shower core on the ice surface, the in-ice emission will have to travel larger distances through the ice, where the propagation velocity is significantly lower compared to air. The arrival time of the in-ice emission will therefore increase. The in-air emission enters the ice much closer to the antenna position. Its arrival time is less affected by the distance of the antenna from the shower core, and depends mostly on the depth of the antenna.

For the antennas closer to the point of impact of the shower core on the ice surface (left columns of Figure~\ref{fig:traces_west_axis_comb} and Figure~\ref{fig:traces_north_axis_comb}), the in-air emission dominates. The Cherenkov angle in air is of the order of $1^{\circ}$, and the in-air emission is therefore strongest close to the shower axis~\cite{Schr_der_2017}. For the antennas further out, at a horizontal distance of $80$~m from the shower axis, the in-ice emission dominates, as these antennas are placed in the Cherenkov ring clearly visible in Figure~\ref{fig:fp_combined}.

For the in-air emission the west component of the electric field is the strongest, showing that the geomagnetic emission dominates. The Askaryan emission is radially polarized and is therefore clearly visible for antennas located on the north axis (Figure~\ref{fig:traces_north_axis_comb}), showing a signal in the north component close to the shower core. This component also shows the expected polarization flip when moving from one side of the shower core to the other. The Askaryan emission can also be seen through its interference with the geomagnetic emission in the antennas on the west axis (Figure~\ref{fig:traces_west_axis_comb}), causing the total signal in the west component to be stronger on the negative side of the axis compared to that on the positive side. The vertical component of the in-air emission is always small since the in-air emission comes from above the ice surface, corresponding to down-going rays toward the given antenna positions and thus horizontally polarized electric fields. It is noticed that the pulses of the west component and the vertical component for the in-air emission at 20~m from the axis center shown in Figure~\ref{fig:traces_west_axis_comb} and Figure~\ref{fig:traces_north_axis_comb} are different in shape. A possible explanation would be the difference in pulse shapes due to the geomagnetic emission driving the east-west component, and the charge-excess emission driving the vertical component~\cite{Scholten_2016}.

For the in-ice emission we see a strong west component for antennas on the west axis (Figure~\ref{fig:traces_west_axis_comb}), and a strong north component for antennas on the north axis (Figure~\ref{fig:traces_north_axis_comb}), showing the radial polarization of the Askaryan emission. The expected polarization flip when moving from one side of the shower core to the other is also clearly visible. In contrast to the in-air emission, the in-ice emission does have significant vertical components. Since the given antenna positions are located closer to the emission points the incoming rays are not completely down going, which results in electric fields with both a horizontal and vertical component.

\subsection{In-ice emission without ray tracing}

To show the effect of the exponential density profile on the cosmic-ray in-ice particle cascade development and its corresponding radio emission, we also simulated the cosmic-ray cascade using an ice layer with constant density and constant index of refraction. The ice density was fixed to the value of $359~$kg$/$m$^{3}$, and the corresponding ice refractive index value was fixed to $n=1.35$, i.e. the value at $z = 0$ for the exponential profile given in Equation~\ref{eq:index_profile}.

\begin{figure}
    \includegraphics[trim={0 1.5cm 0 1cm},width=\linewidth]{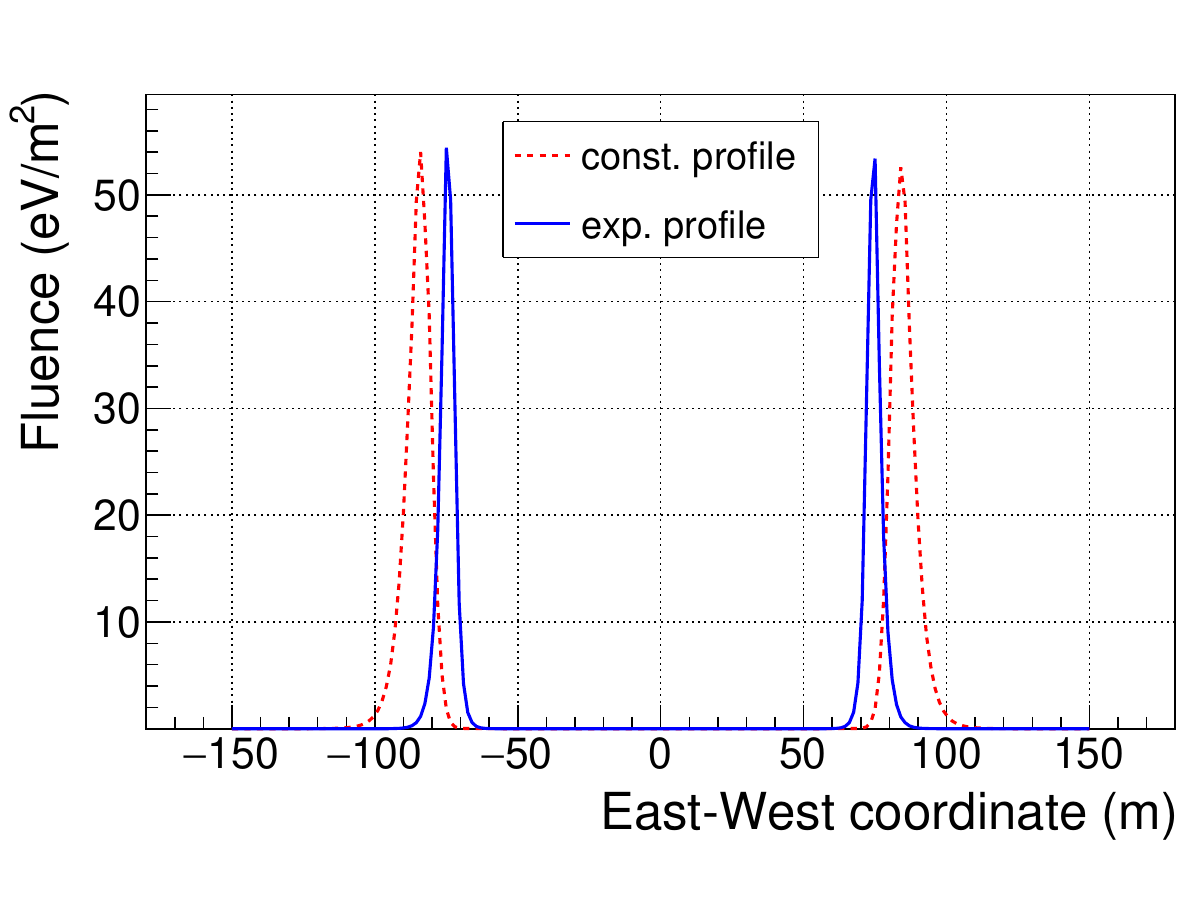}
    \caption{The fluence on the west axis of the simulated cosmic-ray shower at a depth of $100$~m for the in-ice emission only using a constant ice density of $359$~kg/m$^3$ and a constant index of refraction $n = 1.35$, compared to the result using ray tracing with an exponential density and index of refraction profile. In both cases an 8th order digital Butterworth bandpass filter for a frequency band of $300-1000$~MHz was applied. The simulation was performed using 201 antennas using an antenna spacing of 1.5~m, with a primary energy of $E_p = 10^{17}$~eV and zenith angle $\theta = 0$.}
    \label{fig:const_dens_fluence_plot_West_arm}
\end{figure}

Figure~\ref{fig:const_dens_fluence_plot_West_arm} shows the fluence along the west axis for the in-ice emission using the constant density and index of refraction profiles, compared to the result obtained using the exponential profiles and ray tracing. In both cases an 8th order digital Butterworth bandpass filter for a frequency band of $300-1000$~MHz was applied. The simulation was performed using 201 antennas on each axis using an antenna spacing of 1.5~m. As expected, adding ray tracing to the simulation clearly decreases the radius of the Cherenkov ring, as rays bend toward the shower axis when propagating through the ice.

\subsection{Frequency dependence}

\begin{figure*}
    \centering
    \includegraphics[trim={0 0cm 0 0cm}, clip, width=\textwidth]{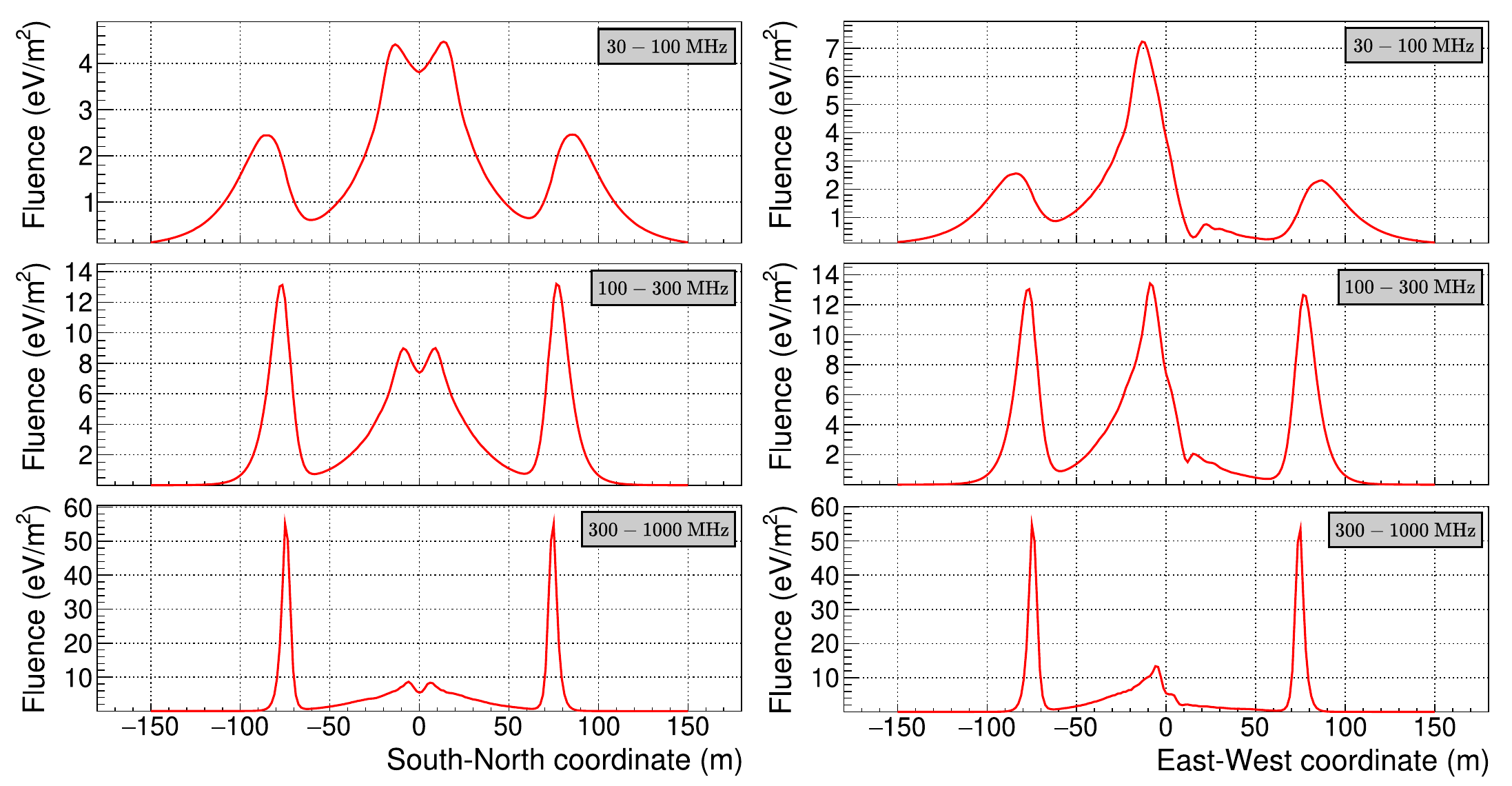}
    \caption{The fluence on the north axis (left) and west axis (right) of the simulated cosmic-ray shower at a depth of 100~m for the combined in-air and in-ice emission when applying an 8th order digital Butterworth bandpass filter, for the frequency bands $30-100$~MHz (upper), $100-300$~MHz (middle) and $300-1000$~MHz (lower). The simulation was performed using 201 antennas each on the north and west axis using an antenna spacing of 1.5~m, with a primary energy of $E_p = 10^{17}$~eV and zenith angle $\theta = 0$.}
    \label{fig:bw_fluence_plot_comb.pdf}
\end{figure*}

To study the frequency dependence of the radio emission we applied an 8th order digital Butterworth bandpass filter to the electric field waveforms for three different frequency bands, and calculated the resulting fluence values for 201 antennas each on the north and west axis with an antenna spacing of 1.5~m, shown in Figure~\ref{fig:bw_fluence_plot_comb.pdf}. 

The two outer fluence peaks correspond to the Cherenkov ring from the in-ice emission, while the inner peak shows the bean shape from the in-air emission. As expected the relative contribution of the in-ice emission at higher frequencies is larger, since due to the higher density, the in-ice particle cascade is more compressed compared to the in-air cascade. In the limit of a pointlike charge, coherence is expected over all frequencies. We also clearly see the width of the Cherenkov cone of the in-ice emission decreasing at higher frequencies, as for higher frequencies the condition for coherence is restricted to smaller length scales and therefore only fulfilled closer to the Cherenkov angle~\cite{ZHS1992}.

\subsection{Primary energy dependence}

\begin{figure*}
    \includegraphics[width=\linewidth]{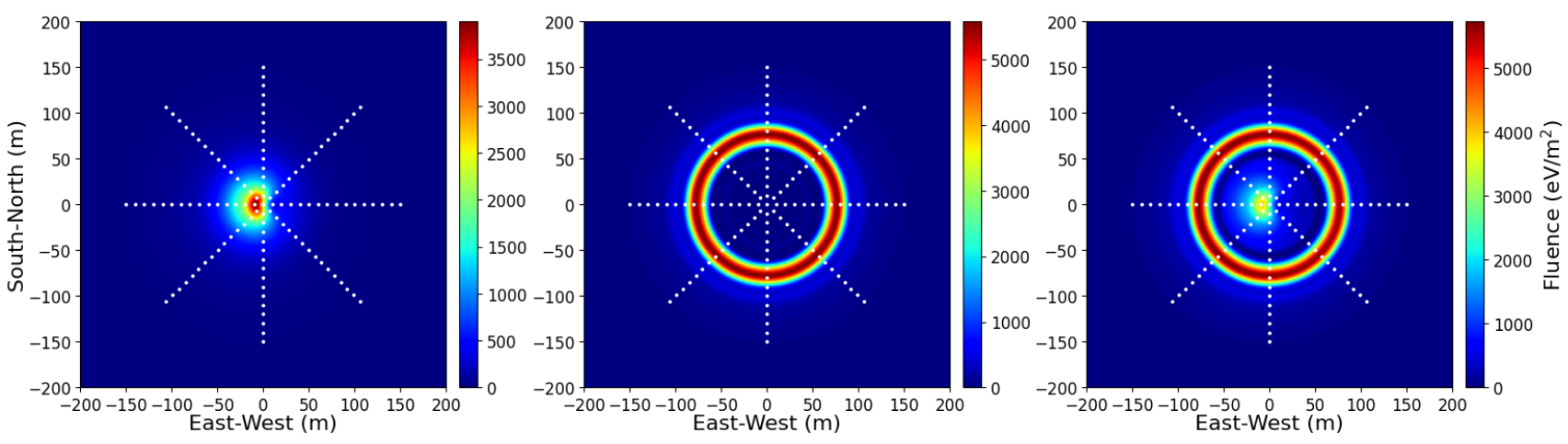}
    \caption{The fluence footprint of the simulated cosmic-ray shower at a depth of 100~m for the in-air emission (left), in-ice emission (middle), and the combined emission (right), using a primary energy of $E_p = 10^{18}$~eV and zenith angle $\theta = 0$. The simulation was performed using 121 antennas in a star-shaped grid with 8 arms and an antenna spacing of 10~m, indicated by the white dots. For the interpolation of the star grid we used the code described in~\cite{Corstanje_Code}.}
    \label{fig:fp_combined_HE}
\end{figure*}

\begin{figure*}
    \centering
    \includegraphics[width=0.9\textwidth]{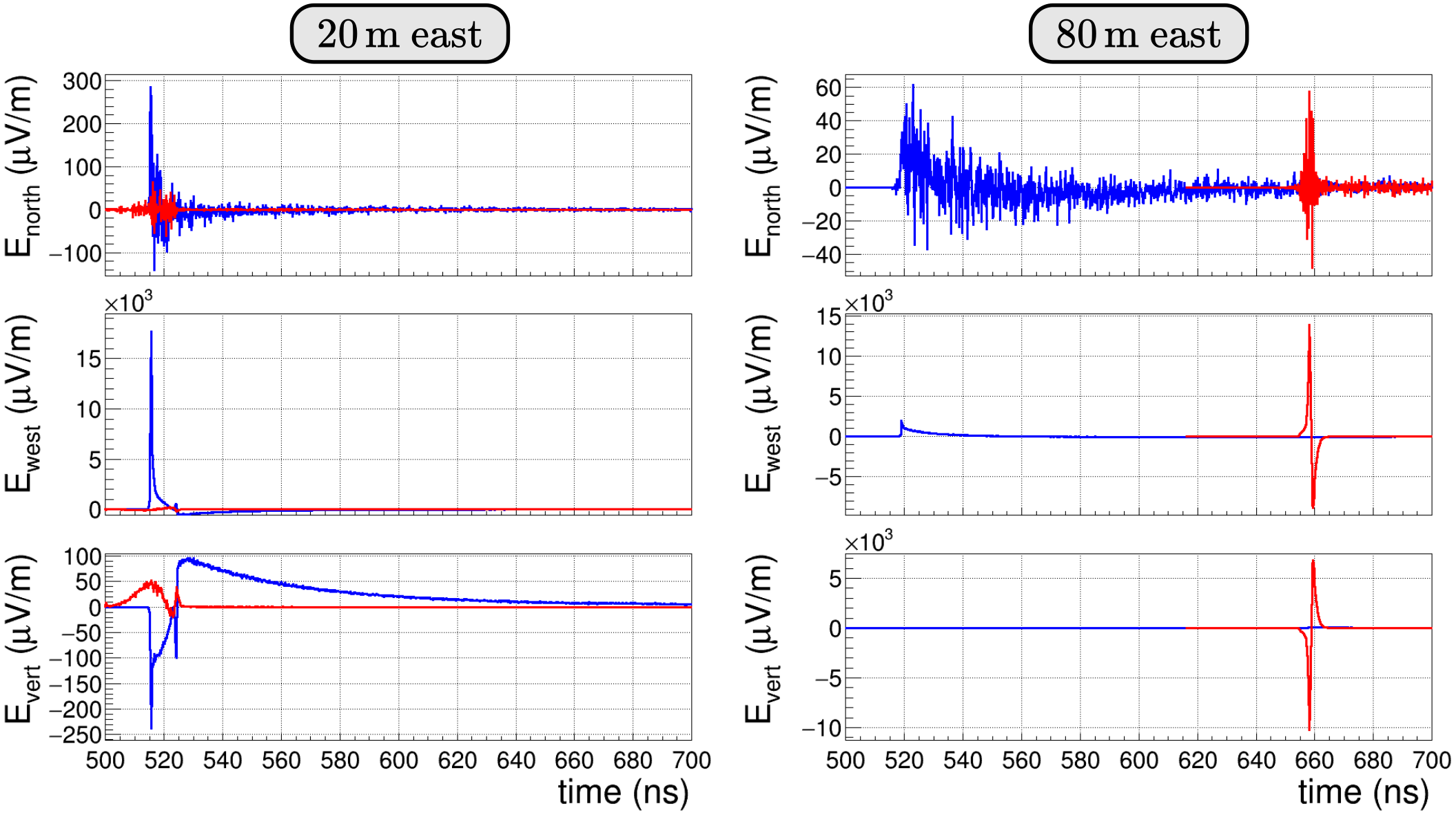}
    \caption{The electric field components as a function of time for 2 different antennas positioned along the west axis at a depth of 100~m, using a primary energy of $E_p = 10^{18}$~eV and zenith angle $\theta = 0$. The coordinate of each antenna is indicated at the top of the plots. The blue line indicates the in-air emission, and the red line indicates the in-ice emission. Note that the range on the y axis is different for every plot.}
    \label{fig:traces_west_axis_comb_HE}
\end{figure*}

Figure~\ref{fig:fp_combined_HE} shows the radio footprint observed by in-ice antennas at $100~$m depth when the primary energy is increased by an order of magnitude to $E_p = 10^{18}$~eV. The longitudinal particle and energy distributions for the in-air particle cascade are shown in Appendix~\ref{App:shower_profiles} (Figure~\ref{fig:part_dist_1e9}), together with the energy deposited in the ice by the in-ice particle cascade (right of Figure~\ref{fig:E_dens_both}). 

The number of particles in the cascade and therefore the amplitude of the electric field scales with primary energy. Following Equation~\ref{Eq:fluence}, increasing the energy of the primary particle by a factor of $10$ should thus lead to an increase in the fluence by a factor of $100$. However, comparing Figure~\ref{fig:fp_combined_HE} to Figure~\ref{fig:fp_combined}, we see that the fluence from the in-ice emission at a primary energy of $10^{18}$~eV has increased significantly more compared to that of the in-air emission. An air shower particle cascade with a higher primary energy will develop deeper into the atmosphere, which means a larger fraction of the primary energy is still contained within the particle shower core when it reaches the ice. Increasing the primary energy of the shower by a factor of $10$ has lead to an increase in the total energy deposited in the ice within a radius of $1$~m of the shower axis by a factor of $15$. The fluence integrated over the footprint in the ice of the in-ice emission has increased by a factor of $240$, which is indeed close to the expected $15^2$. As a larger part of the shower development has been shifted into the ice, the in-air emission has only increased by a factor of $85$. The effect of increasing the primary energy on the energy contained within the particle shower core is illustrated by Figure~\ref{fig:E_dens_both} in Appendix~\ref{App:shower_profiles}.

Figure~\ref{fig:traces_west_axis_comb_HE} shows the electric field components of two antennas positioned along the west axis at a depth of $100$~m using a primary energy of $E_p = 10^{18}$~eV. Comparing these to the two upper panels of Figure~\ref{fig:traces_west_axis_comb} confirms this behavior. Overall the amplitude of the electric field components of the in-air emission has increased by 1 order of magnitude, which is less than the increase of the amplitude for the in-ice emission on the Cherenkov ring.

\subsection{Double pulse signature}

One of the most important purposes of the cosmic-ray cascade simulation is to be able to extract key features of cosmic-ray showers that can be used to identify them in in-ice radio detector data and distinguish them from neutrino-induced cascade radio signals. One such feature is the difference in the time of arrival of the in-air and in-ice emission. This feature causes the two radio pulses to separate out in time and provides us with a double pulse signature that can possibly be used to identify cosmic-ray radio signals in data.

\begin{figure}
    \includegraphics[trim={0cm 1.5cm 0cm 0.5cm},clip,width=\linewidth]{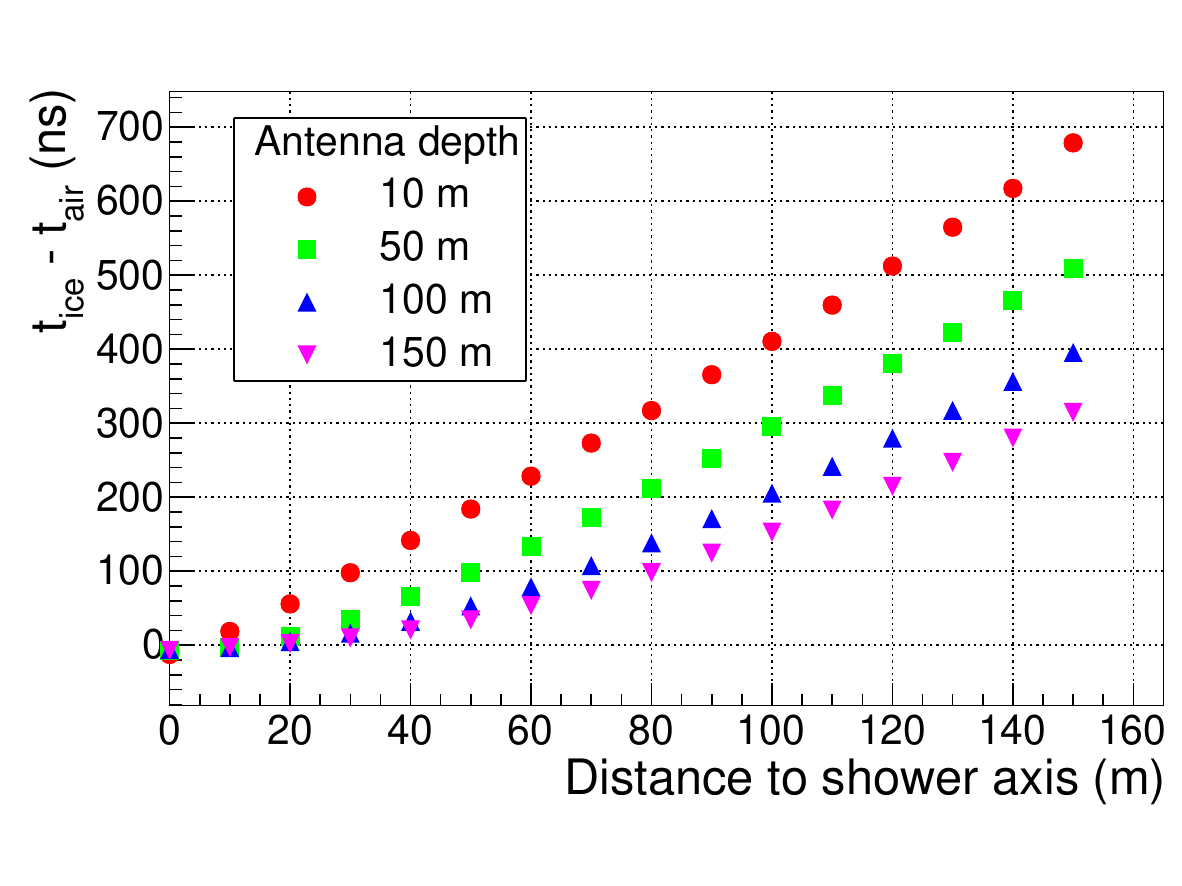}
\caption{ \label{fig:time_diffs} The difference in arrival time of air and ice radio pulses as a function of distance to the shower axis for varying depths as indicated by the legend, using a primary energy $E_p = 10^{17}$~eV and a zenith angle $\theta = 0$. The arrival time was defined as the time where the Hilbert envelope reaches $33\%$ of its maximum value, after passing an 8th order digital Butterworth bandpass filter for the frequency band $30-1000$~MHz, averaged over the three components.}
\end{figure}

Figure~\ref{fig:time_diffs} shows the time delays measured between the air and ice pulses as a function of the distance of the antenna to the shower axis for varying antenna depths, using a primary energy $E_p = 10^{17}$~eV, zenith angle $\theta = 0^{\circ}$ and azimuth angle $\phi = 0^{\circ}$. To determine the arrival time of either an in-air or in-ice pulse we constructed the Hilbert envelope of the given pulse after applying an 8th order digital Butterworth bandpass filter for the frequency band $30-1000$~MHz. The arrival time was then defined as the earliest time where the envelope reaches $33\%$ of its maximum value.

If the receiver is far out from the shower axis and close to the surface, the in-air emission will arrive well before the in-ice emission does. As mentioned earlier the propagation velocity of the radio emission is significantly lower in ice compared to air, delaying the in-ice emission but hardly affecting the in-air emission. If the depth of the receiver increases, the time difference between the two pulses decreases, as now also the in-air emission is affected by the in-ice delay. When the receiver is located on the shower axis, the in-ice and in-air emission will arrive almost simultaneously, with the in-ice emission arriving slightly before the in-air emission. Both emission components will travel the same distance through ice, but the in-ice emission has a small head start because the in-air particle shower moves slightly faster through air than its corresponding in-air radio emission does.

This double pulse feature could be used by in-ice radio experiments to search for cosmic-ray particle cascade signals, along other properties like arrival direction. Especially when the polarization of the pulses is taken into account, it can be a strong discriminator for cosmic-ray events and constrain the event geometry. One of the two pulses should follow the geomagnetic polarization ($\vec{v} \times \vec{B}$), while the other pulse should show radial polarization. Furthermore, as shown by Figure~\ref{fig:time_diffs}, by measuring the time difference between both pulses at a given depth an estimation of the distance between the shower axis and the receiver could be made. It should be noted however that a degeneracy might arise when considering different shower arrival directions.

\section{Conclusion}

We presented FAERIE, a framework developed to simulate the radio emission of cosmic-ray induced particle cascades for observers located in ice. It is the first full Monte-Carlo framework including the shower development and corresponding emission in both air and ice. Ray tracing is applied to account for the changing index of refraction in both media, as well as the transition of the radiation from air to ice. To simulate the in-air particle cascade and its corresponding radio emission, we used the air shower simulation program CORSIKA 7.7500 with a modified version of CoREAS to handle signal propagation to antenna positions in ice. To simulate the propagation of the air shower through ice and its corresponding radio emission, we developed a program based on the GEANT4 simulation toolkit.

Ray tracing was performed analytically by using exponential refractive index profiles for both air and ice, leading to significant ray bending for in-ice radio propagation. For ice-to-ice ray tracing in general two contributions are found, which we called the direct and indirect emission, the latter consisting of reflected and refracted ray solutions. To achieve reasonable computation times the ray tracing was performed through the interpolation of premade tables, which implies that also slower, more accurate ray tracers can be used.

To account for ray tracing we investigated the behavior of the endpoint formalism, and found that it still holds when correctly interpreting the different variables. We also included the transmission Fresnel coefficients for air-to-ice rays moving through the air-ice boundary, and the reflection Fresnel coefficients for ice-to-ice rays reflecting on the ice-air boundary. Furthermore, we included a focusing factor to account for ray convergence and divergence.

We discussed the first simulation results, using a proton with a primary energy of $E_p = 10^{17}$~eV, a zenith angle $\theta = 0^{\circ}$ and an azimuth angle $\phi = 0^{\circ}$. The fluence footprint of the in-air emission is marked by the typical bean-shaped pattern arising from the interference between geomagnetic and Askaryan emission, while the in-ice emission fluence footprint shows a symmetric Cherenkov ring associated to Askaryan emission. The combination of both gives a unique footprint showing the bean-shaped pattern from the in-air emission enclosed by the Cherenkov ring from the in-ice emission. We also showed the electric field as a function of time for different polarizations and antenna positions along two different axes, and discussed general features characterizing the in-air and in-ice components. We compared the in-ice radio emission to the result obtained without ray tracing, applied an 8th order digital Butterworth bandpass filter to study the frequency dependence of the emission, and studied the dependence of the different components of the radio emission on the energy of the primary particle. Finally we discussed the typical double pulse signature of the time traces, arising due to the different arrival times of the in-air and in-ice emission. We showed that in general the in-air emission arrives before the in-ice emission, and that the time difference between both pulses shows a clear relation with distance between receiver and shower axis, as well as receiver depth.

FAERIE combines the CORSIKA 7.7500 and the GEANT4 simulation toolkit for a full Monte-Carlo simulation of the radio emission from cosmic-ray particle cascades, taking into account the propagation and emission in both air and ice. In the future it will be possible to simulate similar cross-media scenarios in one joint step within CORSIKA~8~\cite{CORSIKA8_1, CORSIKA8_2}.

\section{Acknowledgements}

This work was supported by the Flemish Foundation for Scientific Research (FWO-G085820N, FWO-I002119N, and FWO-I001423N) as well as the European Research Council under the European Unions Horizon 2020 research and innovation program (No 805486 - K. D. de Vries). \\

The code is available on request. A request can be made by contacting the authors via e-mail. 

\appendix

\section{Shower profiles}\label{App:shower_profiles}

\begin{figure*}
    \includegraphics[width=0.9\linewidth]{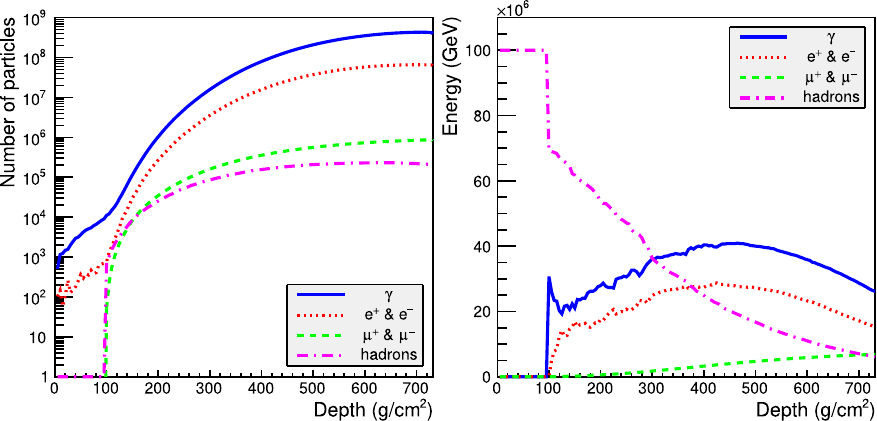}
    \caption{The number of particles (left) and the distribution of the energy (right) in the air shower as a function of depth with primary energy $E_p = 10^{17}$~eV and zenith angle $\theta = 0$. The particle distributions are obtained over the full radial extent of the in-air particle cascade. Similar results were shown in~\cite{DeKockere2022}.}
    \label{fig:part_dist_1e8}
\end{figure*}

\begin{figure*}
    \includegraphics[width=0.9\linewidth]{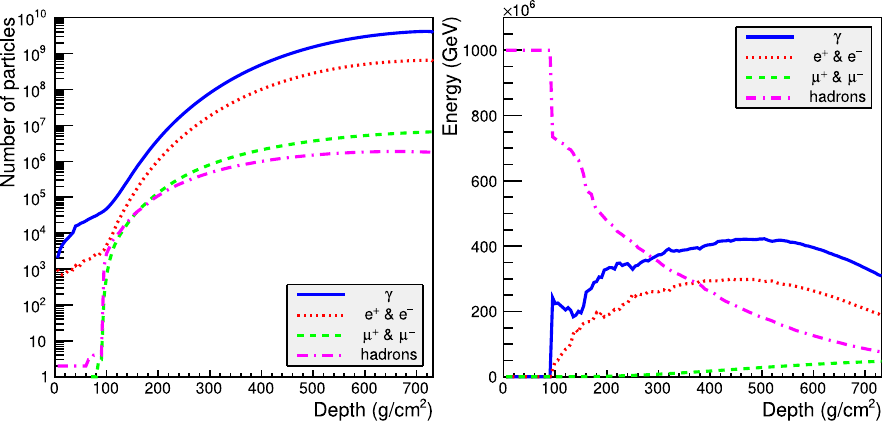}
    \caption{The number of particles (left) and the distribution of the energy (right) in the air shower as a function of depth with primary energy $E_p = 10^{18}$~eV and zenith angle $\theta = 0$. The particle distributions are obtained over the full radial extent of the in-air particle cascade. Similar results were shown in~\cite{DeKockere2022}.}
    \label{fig:part_dist_1e9}
\end{figure*}

\begin{figure*}
    \includegraphics[width=0.95\linewidth]{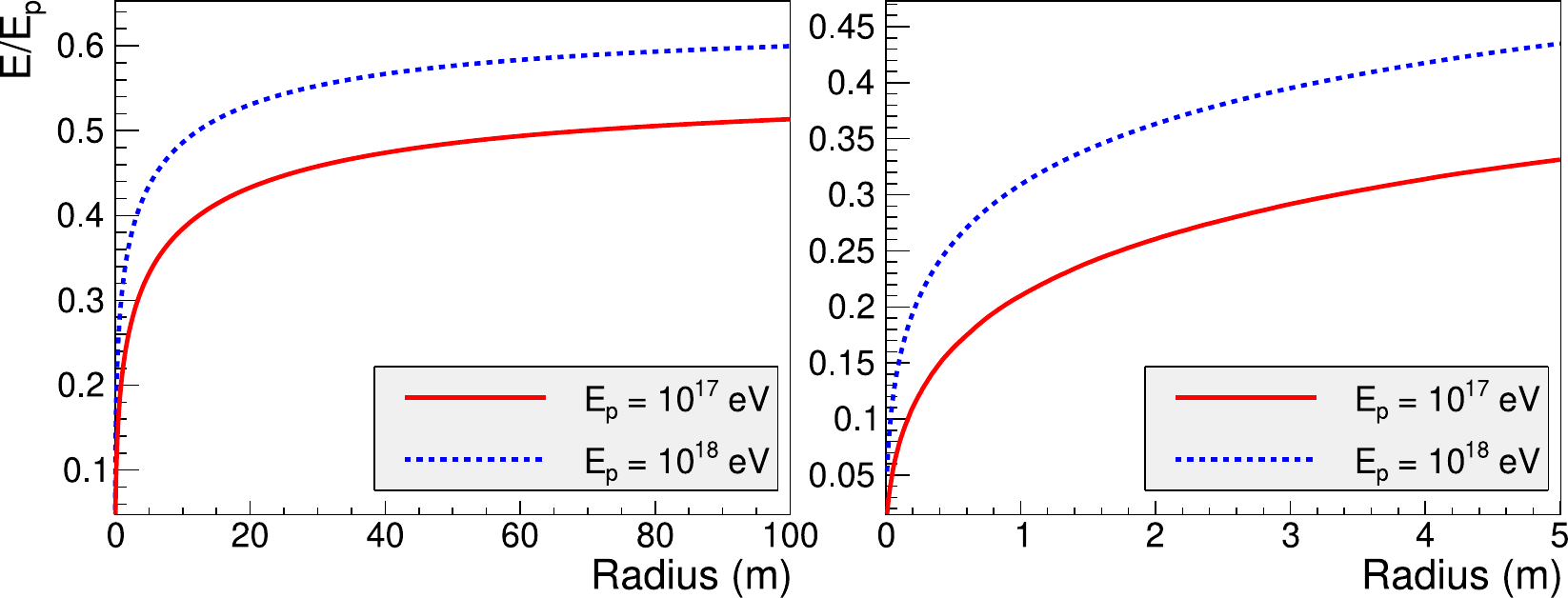}
    \caption{The total energy contained within a given radius from the shower core at the air-ice boundary up to $100$~m (left) and in more detail up to $5$~m (right), for the simulated shower with a primary energy $E_p = 10^{17}$~eV and zenith angle $\theta = 0$ and the simulated shower with a primary energy $E_p = 10^{18}$~eV and zenith angle $\theta = 0$. The energy is expressed as a fraction of the primary energy $E_p$. The air-ice boundary is located at an altitude of $2.835$~km, which corresponds to a vertical atmospheric depth of $729$~g/cm$^2$.}
    \label{fig:Ecum_comb}
\end{figure*}

\begin{figure*}
    \includegraphics[width=\linewidth]{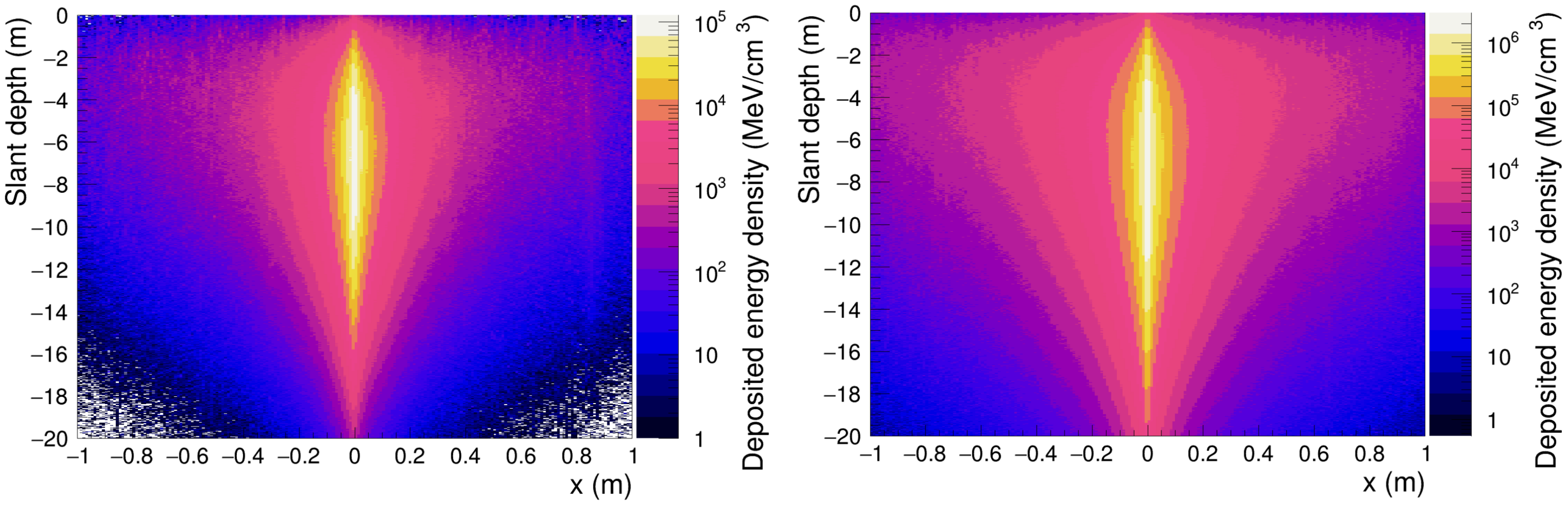}
    \caption{The energy deposited in ice by the in-ice particle cascade for the simulated shower with a primary energy $E_p = 10^{17}$~eV and zenith angle $\theta = 0$ (left) and the simulated shower with a primary energy $E_p = 10^{18}$~eV and zenith angle $\theta = 0$ (right). Shown here is the deposited energy density within a vertical 1-cm wide slice going through the center of the particle shower. The air-ice boundary is located at an altitude of $2.835$~km, which corresponds to a vertical atmospheric depth of $729$~g/cm$^2$. Similar results were shown in~\cite{DeKockere2022}.}
    \label{fig:E_dens_both}
\end{figure*}

Figures~\ref{fig:part_dist_1e8}~and~\ref{fig:part_dist_1e9} show the longitudinal particle and energy distributions of respectively the simulated air shower at a primary energy of $E_p = 10^{17}$~eV and $E_p = 10^{18}$~eV, both with a zenith angle $\theta = 0$. The air-ice boundary is located at a depth of $729$~g/cm$^2$, which means the distributions only show the in-air particle cascade. Figure~\ref{fig:Ecum_comb} shows the total energy contained within a given radius from the shower core at the air-ice boundary for both showers. The energy deposited in the ice by both particle cascades is shown in Figure~\ref{fig:E_dens_both}.

\section{Atmospheric model}\label{App:atmosphere}

The density of the atmosphere is modeled by five consecutive layers. The lower four layers each follow an exponential profile given by
\begin{equation}\label{Eq:atm_dens_profile_a}
    T(h) = a_i + b_i e^{-h/c_i} \quad i = 1, 2, 3, 4,
\end{equation}
with $T$ the mass overburden and $h$ the height~\cite{CORSIKA_manual}. The fifth, topmost layer follows a linear profile given by
\begin{equation}\label{Eq:atm_dens_profile_b}
    T(h) = a_5 - b_5 h / c_5.
\end{equation}
For the simulations described in this work, the parameters used are given in Table~\ref{table:atmosphere}. The refractive index $n$ of the atmosphere is modeled by five consecutive layers as well, all following an exponential profile given by
\begin{equation}\label{Eq:atm_refr_profile}
    n(h) = 1 + B_i e^{-C_i h},
\end{equation}
based on the five atmospheric density layer profiles, as described in~\cite{Latif_thesis_2020}. The parameters for the index of refraction profiles are summarized in Table~\ref{table:atmosphere_n}. 

\setlength{\tabcolsep}{16pt}
\renewcommand{\arraystretch}{1.5}
\begin{table*}[]
\begin{tabular}{|c|c|c|c|c|}
\hline
Layer & Altitude interval (m) & $a_i$ (g/cm$^2$) & $b_i$ (g/cm$^2$) & $c_i$ (cm)  \\ \hline
1     & 0 - 3217              & -1.13352$\times 10^2$     & 1.19439$\times 10^{3}$      & 8.10969$\times 10^{5}$ \\ \hline
2     & 3217 - 8364           & -9.73769     & 1.10328$\times 10^{3}$      & 7.06357$\times 10^{5}$ \\ \hline
3     & 8364 - 23142          & -2.18461$\times 10^{-1}$     & 1.10964$\times 10^{3}$      & 6.86443$\times 10^{5}$ \\ \hline
4     & 23142 - 100000        & 7.95615$\times 10^{-4}$      & 1.12499$\times 10^{3}$      & 6.82494$\times 10^{5}$ \\ \hline
5     & \textgreater 100000   & 1.12829$\times 10^{-2}$      & 1.00000          & 1.00000$\times 10^{9}$ \\ \hline
\end{tabular}
\caption{The numerical values of the parameters in Equations~\ref{Eq:atm_dens_profile_a}~and~\ref{Eq:atm_dens_profile_b} describing the density of the atmosphere used for the simulations in this work.}
\label{table:atmosphere}
\end{table*}

\begin{table*}[]
\begin{tabular}{|c|c|c|c|}
\hline
Layer & Altitude interval (m) & $B_i$                    & $C_i$ (m$^{-1}$)         \\ \hline
1     & 0 - 3217              & 3.28911 $\times 10^{-4}$ & 1.23309 $\times 10^{-4}$ \\ \hline
2     & 3217 - 8364           & 3.48817 $\times 10^{-4}$ & 1.41571 $\times 10^{-4}$ \\ \hline
3     & 8364 - 23142          & 3.61006 $\times 10^{-4}$ & 1.45679 $\times 10^{-4}$ \\ \hline
4     & 23142 - 100000        & 3.68118 $\times 10^{-4}$ & 1.46522 $\times 10^{-4}$ \\ \hline
5     & \textgreater 100000   & 3.68404$\times 10^{-4}$  & 1.46522 $\times 10^{-4}$ \\ \hline
\end{tabular}
\caption{The numerical values of the parameters in Equation~\ref{Eq:atm_refr_profile} describing the refractive index of the atmosphere used for the simulations in this work.}
\label{table:atmosphere_n}
\end{table*}

\section{CPU time and job run time}\label{App:run_time}

\begin{figure}
    \centering
    \includegraphics[trim={0cm 1.5cm 0cm 0.5cm},clip,width=\linewidth]{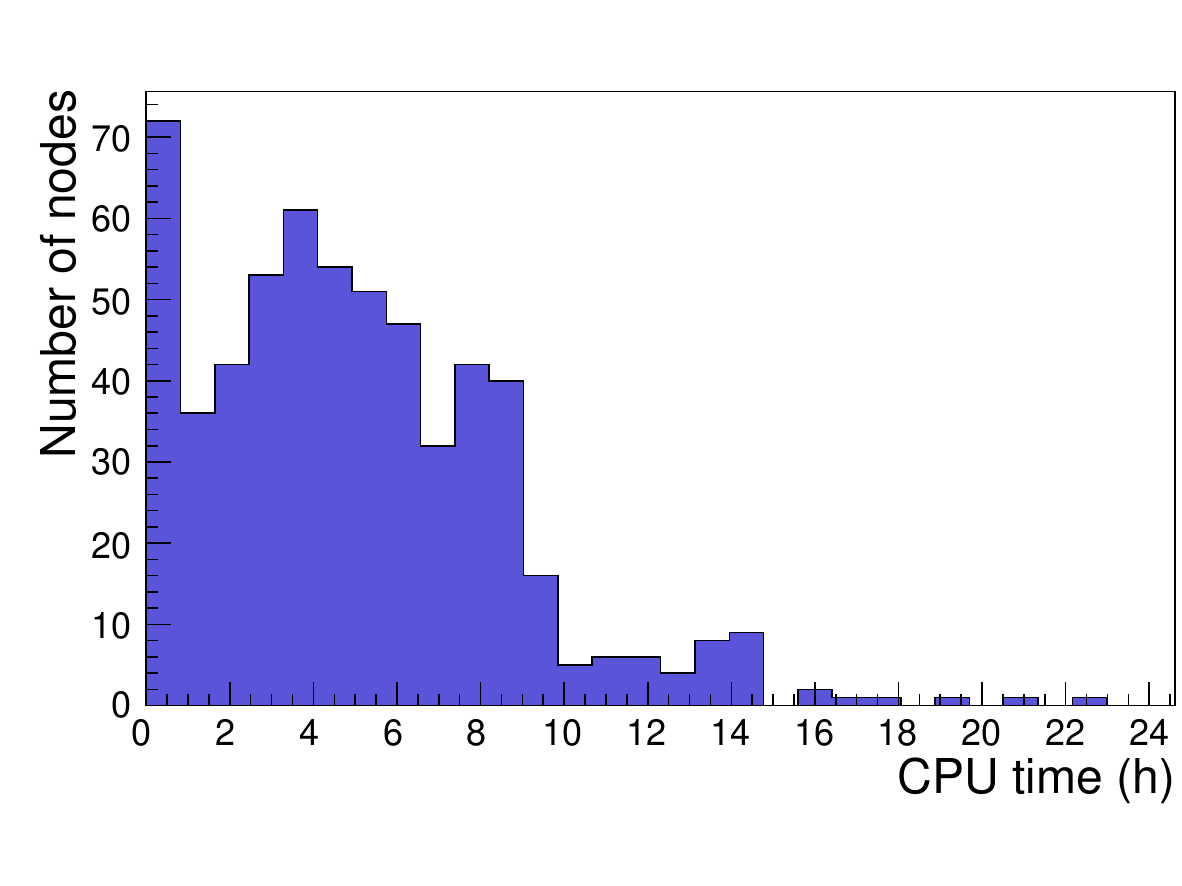}
    \caption{A distribution of the CPU time of 591 cores used for the simulation of the in-ice particle cascade and the corresponding radio emission for 121 antennas in the ice, using a primary energy $E_p = 10^{17}$~eV and zenith angle $\theta = 0$.}
    \label{fig:runtime_histo}
\end{figure}

Running the simulation of a cosmic-ray cascade using FAERIE requires the three following steps: (1) running the in-air simulation with CORSIKA and CoREAS, (2) processing the CORSIKA particle output file to generate input files for the GEANT4-based module, and (3) running the in-ice simulation with the GEANT4-based module. Since multiple CPUs were used during steps (1) and (3) of the simulation process, we can distinguish between CPU time (the total time spent by a single CPU) and the job run time (the time between the start and the end of the simulation process).

To simulate the cosmic-ray cascade initiated by a $E_p = 10^{17}$~eV proton entering the atmosphere at a zenith angle $\theta = 0^{\circ}$ using a star-shaped antenna grid of 121 antennas, the in-air simulation with CORSIKA and CoREAS was split up over 8 different CPUs. Each CPU simulated the full cosmic-ray air shower and its corresponding radio emission for 15 of the 121 antennas, with one of the CPUs including a 16th antenna at (0,0). The total CPU time for the in-air simulation was $171$h, which gives an average CPU time of $21.3$h. The longest CPU time was $26.8$h, which equals the job run time.

To reduce the CPU time of the in-ice simulation, only the particles within a radius of $1$~m of the shower axis are propagated through the ice~\cite{DeKockere2022}. Furthermore we split up the particle footprint into separate parts, running the in-ice simulation for each part on a different CPU. Note that following this approach, a single CPU can simulate the full antenna grid of 121 antennas. The splitting process is based on the energy of the particles, where the amount of particles per part can vary from $540,000$ (energies below $10^7$~eV) down to only 1 (starting at energies of $10^{13}$~eV). Figure~\ref{fig:runtime_histo} shows the distribution of the CPU time of the $591$ cores used for the in-ice simulation. Most of the CPUs manage to stay within $10$~h of CPU time. Some cores however take significantly longer, driving up the job run time to approximately $24$~h. The distribution shows there is still room for optimization in the splitting process. Combining the particle groups corresponding to the simulations with low CPU time would decrease the amount of cores needed, while splitting up the particle groups corresponding to the simulations with high CPU time would decrease the job run time. The longest CPU time for a simulation of the propagation of a single particle was $12.2$~h, which means for this specific case the job run time will always be at least $12.2$~h, unless the antenna grid is distributed over different cores as well. The total CPU time was $2985$~h.

\bibliography{Bibliography}

\providecommand{\noopsort}[1]{}\providecommand{\singleletter}[1]{#1}%
\begin{thebibliography}{63}%
\makeatletter
\providecommand \@ifxundefined [1]{%
 \@ifx{#1\undefined}
}%
\providecommand \@ifnum [1]{%
 \ifnum #1\expandafter \@firstoftwo
 \else \expandafter \@secondoftwo
 \fi
}%
\providecommand \@ifx [1]{%
 \ifx #1\expandafter \@firstoftwo
 \else \expandafter \@secondoftwo
 \fi
}%
\providecommand \natexlab [1]{#1}%
\providecommand \enquote  [1]{``#1''}%
\providecommand \bibnamefont  [1]{#1}%
\providecommand \bibfnamefont [1]{#1}%
\providecommand \citenamefont [1]{#1}%
\providecommand \href@noop [0]{\@secondoftwo}%
\providecommand \href [0]{\begingroup \@sanitize@url \@href}%
\providecommand \@href[1]{\@@startlink{#1}\@@href}%
\providecommand \@@href[1]{\endgroup#1\@@endlink}%
\providecommand \@sanitize@url [0]{\catcode `\\12\catcode `\$12\catcode
  `\&12\catcode `\#12\catcode `\^12\catcode `\_12\catcode `\%12\relax}%
\providecommand \@@startlink[1]{}%
\providecommand \@@endlink[0]{}%
\providecommand \url  [0]{\begingroup\@sanitize@url \@url }%
\providecommand \@url [1]{\endgroup\@href {#1}{\urlprefix }}%
\providecommand \urlprefix  [0]{URL }%
\providecommand \Eprint [0]{\href }%
\providecommand \doibase [0]{https://doi.org/}%
\providecommand \selectlanguage [0]{\@gobble}%
\providecommand \bibinfo  [0]{\@secondoftwo}%
\providecommand \bibfield  [0]{\@secondoftwo}%
\providecommand \translation [1]{[#1]}%
\providecommand \BibitemOpen [0]{}%
\providecommand \bibitemStop [0]{}%
\providecommand \bibitemNoStop [0]{.\EOS\space}%
\providecommand \EOS [0]{\spacefactor3000\relax}%
\providecommand \BibitemShut  [1]{\csname bibitem#1\endcsname}%
\let\auto@bib@innerbib\@empty
\bibitem [{\citenamefont {Aartsen}\ \emph {et~al.}(2013)\citenamefont {Aartsen}
  \emph {et~al.}}]{aatsen_etal2013}%
  \BibitemOpen
  \bibfield  {author} {\bibinfo {author} {\bibfnamefont {M.~G.}\ \bibnamefont
  {Aartsen}} \emph {et~al.} (\bibinfo {collaboration} {IceCube}),\ }\bibfield
  {title} {\bibinfo {title} {{Evidence for High-Energy Extraterrestrial
  Neutrinos at the IceCube Detector}},\ }\href
  {https://doi.org/10.1126/science.1242856} {\bibfield  {journal} {\bibinfo
  {journal} {Science}\ }\textbf {\bibinfo {volume} {342}},\ \bibinfo {pages}
  {1242856} (\bibinfo {year} {2013})},\ \Eprint
  {https://arxiv.org/abs/1311.5238} {arXiv:1311.5238 [astro-ph.HE]}
  \BibitemShut {NoStop}%
\bibitem [{\citenamefont {Aartsen}\ \emph
  {et~al.}(2018{\natexlab{a}})\citenamefont {Aartsen} \emph
  {et~al.}}]{IceCube2018a}%
  \BibitemOpen
  \bibfield  {author} {\bibinfo {author} {\bibfnamefont {M.~G.}\ \bibnamefont
  {Aartsen}} \emph {et~al.},\ }\bibfield  {title} {\bibinfo {title}
  {{Multimessenger observations of a flaring blazar coincident with high-energy
  neutrino IceCube-170922A}},\ }\bibfield  {journal} {\bibinfo  {journal}
  {Science}\ }\textbf {\bibinfo {volume} {361}},\ \href
  {https://doi.org/10.1126/science.aat1378} {10.1126/science.aat1378} (\bibinfo
  {year} {2018}{\natexlab{a}})\BibitemShut {NoStop}%
\bibitem [{\citenamefont {Aartsen}\ \emph
  {et~al.}(2018{\natexlab{b}})\citenamefont {Aartsen} \emph
  {et~al.}}]{IceCube2018b}%
  \BibitemOpen
  \bibfield  {author} {\bibinfo {author} {\bibfnamefont {M.~G.}\ \bibnamefont
  {Aartsen}} \emph {et~al.},\ }\bibfield  {title} {\bibinfo {title} {{Neutrino
  emission from the direction of the blazar TXS 0506+056 prior to the
  IceCube-170922A alert}},\ }\href {https://doi.org/10.1126/science.aat2890}
  {\bibfield  {journal} {\bibinfo  {journal} {Science}\ }\textbf {\bibinfo
  {volume} {361}},\ \bibinfo {pages} {147–151} (\bibinfo {year}
  {2018}{\natexlab{b}})}\BibitemShut {NoStop}%
\bibitem [{\citenamefont {Abbasi}\ \emph {et~al.}(2022)\citenamefont {Abbasi}
  \emph {et~al.}}]{IceCube2022}%
  \BibitemOpen
  \bibfield  {author} {\bibinfo {author} {\bibfnamefont {R.}~\bibnamefont
  {Abbasi}} \emph {et~al.},\ }\bibfield  {title} {\bibinfo {title} {{Evidence
  for neutrino emission from the nearby active galaxy NGC 1068}},\ }\href
  {https://doi.org/10.1126/science.abg3395} {\bibfield  {journal} {\bibinfo
  {journal} {Science}\ }\textbf {\bibinfo {volume} {378}},\ \bibinfo {pages}
  {538–543} (\bibinfo {year} {2022})}\BibitemShut {NoStop}%
\bibitem [{\citenamefont {Allison}\ \emph {et~al.}(2012)\citenamefont
  {Allison}, \citenamefont {Auffenberg}, \citenamefont {Bard}, \citenamefont
  {Beatty}, \citenamefont {Besson}, \citenamefont {Böser} \emph
  {et~al.}}]{ALLISON2012457}%
  \BibitemOpen
  \bibfield  {author} {\bibinfo {author} {\bibfnamefont {P.}~\bibnamefont
  {Allison}}, \bibinfo {author} {\bibfnamefont {J.}~\bibnamefont {Auffenberg}},
  \bibinfo {author} {\bibfnamefont {R.}~\bibnamefont {Bard}}, \bibinfo {author}
  {\bibfnamefont {J.}~\bibnamefont {Beatty}}, \bibinfo {author} {\bibfnamefont
  {D.}~\bibnamefont {Besson}}, \bibinfo {author} {\bibfnamefont
  {S.}~\bibnamefont {Böser}}, \emph {et~al.},\ }\bibfield  {title} {\bibinfo
  {title} {{Design and initial performance of the Askaryan Radio Array
  prototype EeV neutrino detector at the South Pole}},\ }\href
  {https://doi.org/https://doi.org/10.1016/j.astropartphys.2011.11.010}
  {\bibfield  {journal} {\bibinfo  {journal} {Astroparticle Physics}\ }\textbf
  {\bibinfo {volume} {35}},\ \bibinfo {pages} {457} (\bibinfo {year}
  {2012})}\BibitemShut {NoStop}%
\bibitem [{\citenamefont {Barrella}\ \emph {et~al.}(2011)\citenamefont
  {Barrella}, \citenamefont {Barwick},\ and\ \citenamefont
  {Saltzberg}}]{Barrella:2010vs}%
  \BibitemOpen
  \bibfield  {author} {\bibinfo {author} {\bibfnamefont {T.}~\bibnamefont
  {Barrella}}, \bibinfo {author} {\bibfnamefont {S.}~\bibnamefont {Barwick}},\
  and\ \bibinfo {author} {\bibfnamefont {D.}~\bibnamefont {Saltzberg}},\
  }\bibfield  {title} {\bibinfo {title} {{Ross Ice Shelf in situ
  radio-frequency ice attenuation}},\ }\href
  {https://doi.org/10.3189/002214311795306691} {\bibfield  {journal} {\bibinfo
  {journal} {J. Glaciol.}\ }\textbf {\bibinfo {volume} {57}},\ \bibinfo {pages}
  {61} (\bibinfo {year} {2011})},\ \Eprint {https://arxiv.org/abs/1011.0477}
  {arXiv:1011.0477 [astro-ph.IM]} \BibitemShut {NoStop}%
\bibitem [{\citenamefont {Besson}\ \emph {et~al.}(2008)\citenamefont {Besson}
  \emph {et~al.}}]{Besson:2007jja}%
  \BibitemOpen
  \bibfield  {author} {\bibinfo {author} {\bibfnamefont {D.~Z.}\ \bibnamefont
  {Besson}} \emph {et~al.},\ }\bibfield  {title} {\bibinfo {title} {{In situ
  radioglaciological measurements near Taylor Dome, Antarctica and implications
  for UHE neutrino astronomy}},\ }\href
  {https://doi.org/10.1016/j.astropartphys.2007.12.004} {\bibfield  {journal}
  {\bibinfo  {journal} {Astropart. Phys.}\ }\textbf {\bibinfo {volume} {29}},\
  \bibinfo {pages} {130} (\bibinfo {year} {2008})},\ \Eprint
  {https://arxiv.org/abs/astro-ph/0703413} {arXiv:astro-ph/0703413}
  \BibitemShut {NoStop}%
\bibitem [{\citenamefont {Barwick}\ \emph {et~al.}(2005)\citenamefont
  {Barwick}, \citenamefont {Besson}, \citenamefont {Gorham},\ and\
  \citenamefont {Saltzberg}}]{Barwick2005}%
  \BibitemOpen
  \bibfield  {author} {\bibinfo {author} {\bibfnamefont {S.}~\bibnamefont
  {Barwick}}, \bibinfo {author} {\bibfnamefont {D.}~\bibnamefont {Besson}},
  \bibinfo {author} {\bibfnamefont {P.}~\bibnamefont {Gorham}},\ and\ \bibinfo
  {author} {\bibfnamefont {D.}~\bibnamefont {Saltzberg}},\ }\bibfield  {title}
  {\bibinfo {title} {{South Polar in situ radio-frequency ice attenuation}},\
  }\href {https://doi.org/10.3189/172756505781829467} {\bibfield  {journal}
  {\bibinfo  {journal} {Journal of Glaciology}\ }\textbf {\bibinfo {volume}
  {51}},\ \bibinfo {pages} {231} (\bibinfo {year} {2005})}\BibitemShut
  {NoStop}%
\bibitem [{\citenamefont {Avva}\ \emph {et~al.}(2015)\citenamefont {Avva},
  \citenamefont {Kovac}, \citenamefont {Miki}, \citenamefont {Saltzberg},\ and\
  \citenamefont {Vieregg}}]{Avva_2015}%
  \BibitemOpen
  \bibfield  {author} {\bibinfo {author} {\bibfnamefont {J.}~\bibnamefont
  {Avva}}, \bibinfo {author} {\bibfnamefont {J.~M.}\ \bibnamefont {Kovac}},
  \bibinfo {author} {\bibfnamefont {C.}~\bibnamefont {Miki}}, \bibinfo {author}
  {\bibfnamefont {D.}~\bibnamefont {Saltzberg}},\ and\ \bibinfo {author}
  {\bibfnamefont {A.~G.}\ \bibnamefont {Vieregg}},\ }\bibfield  {title}
  {\bibinfo {title} {{An in situ measurement of the radio-frequency attenuation
  in ice at Summit Station, Greenland}},\ }\href
  {https://doi.org/10.3189/2015jog15j057} {\bibfield  {journal} {\bibinfo
  {journal} {Journal of Glaciology}\ }\textbf {\bibinfo {volume} {61}},\
  \bibinfo {pages} {1005–1011} (\bibinfo {year} {2015})}\BibitemShut
  {NoStop}%
\bibitem [{\citenamefont {{P. Allison \textit{et al}}}(2020)}]{ARA_paper}%
  \BibitemOpen
  \bibfield  {author} {\bibinfo {author} {\bibnamefont {{P. Allison \textit{et
  al}}}} (\bibinfo {collaboration} {ARA}),\ }\bibfield  {title} {\bibinfo
  {title} {{Constraints on the diffuse flux of ultrahigh energy neutrinos from
  four years of Askaryan Radio Array data in two stations}},\ }\href
  {https://doi.org/10.1103/PhysRevD.102.043021} {\bibfield  {journal} {\bibinfo
   {journal} {Phys. Rev. D}\ }\textbf {\bibinfo {volume} {102}},\ \bibinfo
  {pages} {043021} (\bibinfo {year} {2020})}\BibitemShut {NoStop}%
\bibitem [{\citenamefont {Anker}\ \emph {et~al.}()\citenamefont {Anker},
  \citenamefont {Barwick}, \citenamefont {Bernhoff}, \citenamefont {Besson},
  \citenamefont {Bingefors}, \citenamefont {García-Fernández} \emph
  {et~al.}}]{ARIANNA_paper}%
  \BibitemOpen
  \bibfield  {author} {\bibinfo {author} {\bibfnamefont {A.}~\bibnamefont
  {Anker}}, \bibinfo {author} {\bibfnamefont {S.}~\bibnamefont {Barwick}},
  \bibinfo {author} {\bibfnamefont {H.}~\bibnamefont {Bernhoff}}, \bibinfo
  {author} {\bibfnamefont {D.}~\bibnamefont {Besson}}, \bibinfo {author}
  {\bibfnamefont {N.}~\bibnamefont {Bingefors}}, \bibinfo {author}
  {\bibfnamefont {D.}~\bibnamefont {García-Fernández}}, \emph {et~al.},\
  }\bibfield  {title} {\bibinfo {title} {Probing the angular and polarization
  reconstruction of the arianna detector at the south pole},\ }\href
  {https://doi.org/10.1088/1748-0221/15/09/P09039} {\bibfield  {journal}
  {\bibinfo  {journal} {Journal of Instrumentation}\ }\textbf {\bibinfo
  {volume} {15}},\ \bibinfo {pages} {P09039 (2020)}}\BibitemShut {NoStop}%
\bibitem [{\citenamefont {Aguilar}\ \emph {et~al.}()\citenamefont {Aguilar},
  \citenamefont {Allison}, \citenamefont {Beatty}, \citenamefont {Bernhoff},
  \citenamefont {Besson}, \citenamefont {Bingefors} \emph
  {et~al.}}]{RNO_paper}%
  \BibitemOpen
  \bibfield  {author} {\bibinfo {author} {\bibfnamefont {J.}~\bibnamefont
  {Aguilar}}, \bibinfo {author} {\bibfnamefont {P.}~\bibnamefont {Allison}},
  \bibinfo {author} {\bibfnamefont {J.}~\bibnamefont {Beatty}}, \bibinfo
  {author} {\bibfnamefont {H.}~\bibnamefont {Bernhoff}}, \bibinfo {author}
  {\bibfnamefont {D.}~\bibnamefont {Besson}}, \bibinfo {author} {\bibfnamefont
  {N.}~\bibnamefont {Bingefors}}, \emph {et~al.},\ }\bibfield  {title}
  {\bibinfo {title} {Design and sensitivity of the radio neutrino observatory
  in greenland (rno-g)},\ }\href
  {https://doi.org/10.1088/1748-0221/16/03/P03025} {\bibfield  {journal}
  {\bibinfo  {journal} {Journal of Instrumentation}\ }\textbf {\bibinfo
  {volume} {16}}\bibinfo  {number} { (03)},\ \bibinfo {pages} {P03025
  (2021)}}\BibitemShut {NoStop}%
\bibitem [{\citenamefont {Askaryan}(1962)}]{askaryan1962}%
  \BibitemOpen
\bibfield  {number} {  }\bibfield  {author} {\bibinfo {author} {\bibfnamefont
  {G.~A.}\ \bibnamefont {Askaryan}},\ }\bibfield  {title} {\bibinfo {title}
  {{Excess Negative Charge of an Electron-photon Shower and its Coherent Radio
  Emission}},\ }\href@noop {} {\bibfield  {journal} {\bibinfo  {journal} {Sov.
  Phys. JETP}\ }\textbf {\bibinfo {volume} {14}},\ \bibinfo {pages} {441}
  (\bibinfo {year} {1962})}\BibitemShut {NoStop}%
\bibitem [{\citenamefont {Zas}\ \emph {et~al.}(1992)\citenamefont {Zas},
  \citenamefont {Halzen},\ and\ \citenamefont {Stanev}}]{ZHS1992}%
  \BibitemOpen
  \bibfield  {author} {\bibinfo {author} {\bibfnamefont {E.}~\bibnamefont
  {Zas}}, \bibinfo {author} {\bibfnamefont {F.}~\bibnamefont {Halzen}},\ and\
  \bibinfo {author} {\bibfnamefont {T.}~\bibnamefont {Stanev}},\ }\bibfield
  {title} {\bibinfo {title} {{Electromagnetic pulses from high-energy showers:
  Implications for neutrino detection}},\ }\href
  {https://doi.org/10.1103/PhysRevD.45.362} {\bibfield  {journal} {\bibinfo
  {journal} {Phys. Rev. D}\ }\textbf {\bibinfo {volume} {45}},\ \bibinfo
  {pages} {362} (\bibinfo {year} {1992})}\BibitemShut {NoStop}%
\bibitem [{\citenamefont {Alvarez-Muniz}\ and\ \citenamefont
  {Zas}(1997)}]{Alvarez1997}%
  \BibitemOpen
  \bibfield  {author} {\bibinfo {author} {\bibfnamefont {J.}~\bibnamefont
  {Alvarez-Muniz}}\ and\ \bibinfo {author} {\bibfnamefont {E.}~\bibnamefont
  {Zas}},\ }\bibfield  {title} {\bibinfo {title} {{Cherenkov radio pulses from
  EeV neutrino interactions: the LPM effect}},\ }\href@noop {} {\bibfield
  {journal} {\bibinfo  {journal} {Phys. Lett. B}\ }\textbf {\bibinfo {volume}
  {411}},\ \bibinfo {pages} {218} (\bibinfo {year} {1997})},\ \Eprint
  {https://arxiv.org/abs/arXiv:astro-ph/9706064} {arXiv:astro-ph/9706064}
  \BibitemShut {NoStop}%
\bibitem [{\citenamefont {{Prohira, S. and de Vries, K.D. and Allison, P. and
  Beatty, J. and Besson, D. and Connolly, A. and others}}(2020)}]{RET_Obs}%
  \BibitemOpen
  \bibfield  {author} {\bibinfo {author} {\bibnamefont {{Prohira, S. and de
  Vries, K.D. and Allison, P. and Beatty, J. and Besson, D. and Connolly, A.
  and others}}},\ }\bibfield  {title} {\bibinfo {title} {{Observation of Radar
  Echoes from High-Energy Particle Cascades}},\ }\href
  {https://doi.org/10.1103/PhysRevLett.124.091101} {\bibfield  {journal}
  {\bibinfo  {journal} {Phys. Rev. L}\ }\textbf {\bibinfo {volume} {124}},\
  \bibinfo {eid} {091101} (\bibinfo {year} {2020})},\ \Eprint
  {https://arxiv.org/abs/1910.12830} {arXiv:1910.12830 [astro-ph.HE]}
  \BibitemShut {NoStop}%
\bibitem [{\citenamefont {{Prohira, S. and de Vries, K.D. and Allison, P. and
  Beatty, J. and Besson, D. and Connolly, A. and others}}(2021)}]{RET_CR}%
  \BibitemOpen
  \bibfield  {author} {\bibinfo {author} {\bibnamefont {{Prohira, S. and de
  Vries, K.D. and Allison, P. and Beatty, J. and Besson, D. and Connolly, A.
  and others}}},\ }\bibfield  {title} {\bibinfo {title} {{The Radar Echo
  Telescope for Cosmic Rays: Pathfinder experiment for a next-generation
  neutrino observatory}},\ }\href {https://doi.org/10.1103/PhysRevD.104.102006}
  {\bibfield  {journal} {\bibinfo  {journal} {Phys. Rev. D}\ }\textbf {\bibinfo
  {volume} {104}},\ \bibinfo {eid} {102006} (\bibinfo {year} {2021})},\ \Eprint
  {https://arxiv.org/abs/2104.00459} {arXiv:2104.00459 [astro-ph.IM]}
  \BibitemShut {NoStop}%
\bibitem [{\citenamefont {Gorham}\ \emph {et~al.}(2009)\citenamefont {Gorham},
  \citenamefont {Allison}, \citenamefont {Barwick}, \citenamefont {Beatty},
  \citenamefont {Besson}, \citenamefont {Binns} \emph {et~al.}}]{ANITA}%
  \BibitemOpen
  \bibfield  {author} {\bibinfo {author} {\bibfnamefont {P.}~\bibnamefont
  {Gorham}}, \bibinfo {author} {\bibfnamefont {P.}~\bibnamefont {Allison}},
  \bibinfo {author} {\bibfnamefont {S.}~\bibnamefont {Barwick}}, \bibinfo
  {author} {\bibfnamefont {J.}~\bibnamefont {Beatty}}, \bibinfo {author}
  {\bibfnamefont {D.}~\bibnamefont {Besson}}, \bibinfo {author} {\bibfnamefont
  {W.}~\bibnamefont {Binns}}, \emph {et~al.},\ }\bibfield  {title} {\bibinfo
  {title} {{The Antarctic Impulsive Transient Antenna ultra-high energy
  neutrino detector: Design, performance, and sensitivity for the 2006–2007
  balloon flight}},\ }\href
  {https://doi.org/10.1016/j.astropartphys.2009.05.003} {\bibfield  {journal}
  {\bibinfo  {journal} {Astroparticle Physics}\ }\textbf {\bibinfo {volume}
  {32}},\ \bibinfo {pages} {10–41} (\bibinfo {year} {2009})}\BibitemShut
  {NoStop}%
\bibitem [{\citenamefont {Abarr}\ \emph {et~al.}()\citenamefont {Abarr},
  \citenamefont {Allison}, \citenamefont {Ammerman~Yebra}, \citenamefont
  {Alvarez-Muñiz}, \citenamefont {Beatty}, \citenamefont {Besson} \emph
  {et~al.}}]{PUEO}%
  \BibitemOpen
  \bibfield  {author} {\bibinfo {author} {\bibfnamefont {Q.}~\bibnamefont
  {Abarr}}, \bibinfo {author} {\bibfnamefont {P.}~\bibnamefont {Allison}},
  \bibinfo {author} {\bibfnamefont {J.}~\bibnamefont {Ammerman~Yebra}},
  \bibinfo {author} {\bibfnamefont {J.}~\bibnamefont {Alvarez-Muñiz}},
  \bibinfo {author} {\bibfnamefont {J.}~\bibnamefont {Beatty}}, \bibinfo
  {author} {\bibfnamefont {D.}~\bibnamefont {Besson}}, \emph {et~al.},\
  }\bibfield  {title} {\bibinfo {title} {{The Payload for Ultrahigh Energy
  Observations (PUEO): a white paper}},\ }\href
  {https://doi.org/10.1088/1748-0221/16/08/p08035} {\bibfield  {journal}
  {\bibinfo  {journal} {Journal of Instrumentation}\ }\textbf {\bibinfo
  {volume} {16}}\bibinfo  {number} { (08)},\ \bibinfo {pages} {P08035
  (2021)}}\BibitemShut {NoStop}%
\bibitem [{\citenamefont {Álvarez Muñiz}\ \emph {et~al.}(2019)\citenamefont
  {Álvarez Muñiz}, \citenamefont {Alves~Batista}, \citenamefont
  {Balagopal~V.}, \citenamefont {Bolmont}, \citenamefont {Bustamante},
  \citenamefont {Carvalho} \emph {et~al.}}]{GRAND}%
  \BibitemOpen
\bibfield  {number} {  }\bibfield  {author} {\bibinfo {author} {\bibfnamefont
  {J.}~\bibnamefont {Álvarez Muñiz}}, \bibinfo {author} {\bibfnamefont
  {R.}~\bibnamefont {Alves~Batista}}, \bibinfo {author} {\bibfnamefont
  {A.}~\bibnamefont {Balagopal~V.}}, \bibinfo {author} {\bibfnamefont
  {J.}~\bibnamefont {Bolmont}}, \bibinfo {author} {\bibfnamefont
  {M.}~\bibnamefont {Bustamante}}, \bibinfo {author} {\bibfnamefont
  {W.}~\bibnamefont {Carvalho}}, \emph {et~al.},\ }\bibfield  {title} {\bibinfo
  {title} {{The Giant Radio Array for Neutrino Detection (GRAND): Science and
  design}},\ }\bibfield  {journal} {\bibinfo  {journal} {Science China Physics,
  Mechanics \& Astronomy}\ }\textbf {\bibinfo {volume} {63}},\ \href
  {https://doi.org/10.1007/s11433-018-9385-7} {10.1007/s11433-018-9385-7}
  (\bibinfo {year} {2019})\BibitemShut {NoStop}%
\bibitem [{\citenamefont {Southall}\ \emph {et~al.}(2023)\citenamefont
  {Southall}, \citenamefont {Deaconu}, \citenamefont {Decoene}, \citenamefont
  {Oberla}, \citenamefont {Zeolla}, \citenamefont {Alvarez-Muñiz} \emph
  {et~al.}}]{BEACON}%
  \BibitemOpen
  \bibfield  {author} {\bibinfo {author} {\bibfnamefont {D.}~\bibnamefont
  {Southall}}, \bibinfo {author} {\bibfnamefont {C.}~\bibnamefont {Deaconu}},
  \bibinfo {author} {\bibfnamefont {V.}~\bibnamefont {Decoene}}, \bibinfo
  {author} {\bibfnamefont {E.}~\bibnamefont {Oberla}}, \bibinfo {author}
  {\bibfnamefont {A.}~\bibnamefont {Zeolla}}, \bibinfo {author} {\bibfnamefont
  {J.}~\bibnamefont {Alvarez-Muñiz}}, \emph {et~al.},\ }\bibfield  {title}
  {\bibinfo {title} {{Design and initial performance of the prototype for the
  BEACON instrument for detection of ultrahigh energy particles}},\ }\href
  {https://doi.org/10.1016/j.nima.2022.167889} {\bibfield  {journal} {\bibinfo
  {journal} {Nucl. Instrum. Methods Phys. Res., Sect. A}\ }\textbf {\bibinfo
  {volume} {1048}},\ \bibinfo {pages} {167889} (\bibinfo {year}
  {2023})}\BibitemShut {NoStop}%
\bibitem [{\citenamefont {De~Kockere}\ \emph {et~al.}(2022)\citenamefont
  {De~Kockere}, \citenamefont {de~Vries}, \citenamefont {van Eijndhoven},\ and\
  \citenamefont {Latif}}]{DeKockere2022}%
  \BibitemOpen
  \bibfield  {author} {\bibinfo {author} {\bibfnamefont {S.}~\bibnamefont
  {De~Kockere}}, \bibinfo {author} {\bibfnamefont {K.~D.}\ \bibnamefont
  {de~Vries}}, \bibinfo {author} {\bibfnamefont {N.}~\bibnamefont {van
  Eijndhoven}},\ and\ \bibinfo {author} {\bibfnamefont {U.~A.}\ \bibnamefont
  {Latif}},\ }\bibfield  {title} {\bibinfo {title} {{Simulation of in-ice
  cosmic ray air shower induced particle cascades}},\ }\href
  {https://doi.org/10.1103/PhysRevD.106.043023} {\bibfield  {journal} {\bibinfo
   {journal} {Phys. Rev. D}\ }\textbf {\bibinfo {volume} {106}},\ \bibinfo
  {eid} {043023} (\bibinfo {year} {2022})}\BibitemShut {NoStop}%
\bibitem [{\citenamefont {Heck}\ \emph {et~al.}(1998)\citenamefont {Heck},
  \citenamefont {Knapp}, \citenamefont {Capdevielle}, \citenamefont {Schatz},\
  and\ \citenamefont {Thouw}}]{corsika}%
  \BibitemOpen
  \bibfield  {author} {\bibinfo {author} {\bibfnamefont {D.}~\bibnamefont
  {Heck}}, \bibinfo {author} {\bibfnamefont {J.}~\bibnamefont {Knapp}},
  \bibinfo {author} {\bibfnamefont {J.~N.}\ \bibnamefont {Capdevielle}},
  \bibinfo {author} {\bibfnamefont {G.}~\bibnamefont {Schatz}},\ and\ \bibinfo
  {author} {\bibfnamefont {T.}~\bibnamefont {Thouw}},\ }\bibfield  {title}
  {\bibinfo {title} {{CORSIKA: A Monte Carlo code to simulate extensive air
  showers}},\ }\href@noop {} {\bibfield  {journal} {\bibinfo  {journal} {FZKA}\
  }\textbf {\bibinfo {volume} {6019}} (\bibinfo {year} {1998})}\BibitemShut
  {NoStop}%
\bibitem [{\citenamefont {{J. Allison \textit{et al}}}(2016)}]{Geant4}%
  \BibitemOpen
  \bibfield  {author} {\bibinfo {author} {\bibnamefont {{J. Allison \textit{et
  al}}}},\ }\bibfield  {title} {\bibinfo {title} {{Recent developments in
  Geant4}},\ }\href
  {https://doi.org/https://doi.org/10.1016/j.nima.2016.06.125} {\bibfield
  {journal} {\bibinfo  {journal} {Nucl. Instrum. Methods Phys. Res., Sect. A}\
  }\textbf {\bibinfo {volume} {835}},\ \bibinfo {pages} {186} (\bibinfo {year}
  {2016})}\BibitemShut {NoStop}%
\bibitem [{\citenamefont {James}\ \emph {et~al.}(2011)\citenamefont {James},
  \citenamefont {Falcke}, \citenamefont {Huege},\ and\ \citenamefont
  {Ludwig}}]{corsika_endpoint}%
  \BibitemOpen
  \bibfield  {author} {\bibinfo {author} {\bibfnamefont {C.~W.}\ \bibnamefont
  {James}}, \bibinfo {author} {\bibfnamefont {H.}~\bibnamefont {Falcke}},
  \bibinfo {author} {\bibfnamefont {T.}~\bibnamefont {Huege}},\ and\ \bibinfo
  {author} {\bibfnamefont {M.}~\bibnamefont {Ludwig}},\ }\bibfield  {title}
  {\bibinfo {title} {{General description of electromagnetic radiation
  processes based on instantaneous charge acceleration in ``endpoints''}},\
  }\href {https://doi.org/10.1103/PhysRevE.84.056602} {\bibfield  {journal}
  {\bibinfo  {journal} {Phys. Rev. E}\ }\textbf {\bibinfo {volume} {84}},\
  \bibinfo {pages} {056602} (\bibinfo {year} {2011})}\BibitemShut {NoStop}%
\bibitem [{\citenamefont {Razzaque}\ \emph {et~al.}(2002)\citenamefont
  {Razzaque}, \citenamefont {Seunarine}, \citenamefont {Besson}, \citenamefont
  {McKay}, \citenamefont {Ralston},\ and\ \citenamefont
  {Seckel}}]{Razzaque2002}%
  \BibitemOpen
  \bibfield  {author} {\bibinfo {author} {\bibfnamefont {S.}~\bibnamefont
  {Razzaque}}, \bibinfo {author} {\bibfnamefont {S.}~\bibnamefont {Seunarine}},
  \bibinfo {author} {\bibfnamefont {D.~Z.}\ \bibnamefont {Besson}}, \bibinfo
  {author} {\bibfnamefont {D.~W.}\ \bibnamefont {McKay}}, \bibinfo {author}
  {\bibfnamefont {J.~P.}\ \bibnamefont {Ralston}},\ and\ \bibinfo {author}
  {\bibfnamefont {D.}~\bibnamefont {Seckel}},\ }\bibfield  {title} {\bibinfo
  {title} {{Coherent Radio Pulses From GEANT Generated Electromagnetic Showers
  In Ice}},\ }\href@noop {} {\bibfield  {journal} {\bibinfo  {journal} {Phys.
  Rev. D}\ }\textbf {\bibinfo {volume} {65}},\ \bibinfo {pages} {103002}
  (\bibinfo {year} {2002})},\ \Eprint
  {https://arxiv.org/abs/arXiv:astro-ph/0112505} {arXiv:astro-ph/0112505}
  \BibitemShut {NoStop}%
\bibitem [{\citenamefont {{S. Razzaque, S. Seunarine, S.W. Chambers, D.Z.
  Besson, D.W. McKay, J.P. Ralston and D. Seckel}}(2004)}]{Razzaque2002_add}%
  \BibitemOpen
  \bibfield  {author} {\bibinfo {author} {\bibnamefont {{S. Razzaque, S.
  Seunarine, S.W. Chambers, D.Z. Besson, D.W. McKay, J.P. Ralston and D.
  Seckel}}},\ }\bibfield  {title} {\bibinfo {title} {{Addendum to "Coherent
  radio pulses from GEANT generated electromagnetic showers in ice"}},\
  }\href@noop {} {\bibfield  {journal} {\bibinfo  {journal} {Phys. Rev. D}\
  }\textbf {\bibinfo {volume} {69}},\ \bibinfo {pages} {047101} (\bibinfo
  {year} {2004})},\ \Eprint {https://arxiv.org/abs/arXiv:astro-ph/0306291}
  {arXiv:astro-ph/0306291} \BibitemShut {NoStop}%
\bibitem [{\citenamefont {Bevan}\ \emph {et~al.}(2007)\citenamefont {Bevan},
  \citenamefont {Danaher}, \citenamefont {Perkin}, \citenamefont {Ralph},
  \citenamefont {Rhodes}, \citenamefont {Thompson}, \citenamefont {Sloan},\
  and\ \citenamefont {Waters}}]{Bevan_2007}%
  \BibitemOpen
  \bibfield  {author} {\bibinfo {author} {\bibfnamefont {S.}~\bibnamefont
  {Bevan}}, \bibinfo {author} {\bibfnamefont {S.}~\bibnamefont {Danaher}},
  \bibinfo {author} {\bibfnamefont {J.}~\bibnamefont {Perkin}}, \bibinfo
  {author} {\bibfnamefont {S.}~\bibnamefont {Ralph}}, \bibinfo {author}
  {\bibfnamefont {C.}~\bibnamefont {Rhodes}}, \bibinfo {author} {\bibfnamefont
  {L.}~\bibnamefont {Thompson}}, \bibinfo {author} {\bibfnamefont
  {T.}~\bibnamefont {Sloan}},\ and\ \bibinfo {author} {\bibfnamefont
  {D.}~\bibnamefont {Waters}},\ }\bibfield  {title} {\bibinfo {title}
  {Simulation of ultra high energy neutrino induced showers in ice and water},\
  }\href {https://doi.org/10.1016/j.astropartphys.2007.08.001} {\bibfield
  {journal} {\bibinfo  {journal} {Astroparticle Physics}\ }\textbf {\bibinfo
  {volume} {28}},\ \bibinfo {pages} {366–379} (\bibinfo {year}
  {2007})}\BibitemShut {NoStop}%
\bibitem [{\citenamefont {Seckel}\ \emph {et~al.}(2008)\citenamefont {Seckel},
  \citenamefont {Seunarine}, \citenamefont {Clem},\ and\ \citenamefont
  {Javaid}}]{Seckel2008}%
  \BibitemOpen
  \bibfield  {author} {\bibinfo {author} {\bibfnamefont {D.}~\bibnamefont
  {Seckel}}, \bibinfo {author} {\bibfnamefont {S.}~\bibnamefont {Seunarine}},
  \bibinfo {author} {\bibfnamefont {J.}~\bibnamefont {Clem}},\ and\ \bibinfo
  {author} {\bibfnamefont {A.}~\bibnamefont {Javaid}},\ }\bibfield  {title}
  {\bibinfo {title} {{In-Ice radio detection of air shower cores}},\
  }\href@noop {} {\bibfield  {journal} {\bibinfo  {journal} {Proc. 30th ICRC,
  Ciudad de Mexico: Universidad Nacional Autonoma de Mexico, Merida (Mexico)}\
  }\textbf {\bibinfo {volume} {5}},\ \bibinfo {pages} {1029} (\bibinfo {year}
  {2008})}\BibitemShut {NoStop}%
\bibitem [{\citenamefont {Tueros}\ and\ \citenamefont
  {Sciutto}(2010)}]{Tueros2010}%
  \BibitemOpen
  \bibfield  {author} {\bibinfo {author} {\bibfnamefont {M.}~\bibnamefont
  {Tueros}}\ and\ \bibinfo {author} {\bibfnamefont {S.}~\bibnamefont
  {Sciutto}},\ }\bibfield  {title} {\bibinfo {title} {{TIERRAS: A package to
  simulate high energy cosmic ray showers underground, underwater and
  under-ice}},\ }\href@noop {} {\bibfield  {journal} {\bibinfo  {journal}
  {Computer Physics Communications}\ }\textbf {\bibinfo {volume} {181}},\
  \bibinfo {pages} {380} (\bibinfo {year} {2010})}\BibitemShut {NoStop}%
\bibitem [{\citenamefont {Alvarez-Muñiz}\ \emph {et~al.}(2012)\citenamefont
  {Alvarez-Muñiz}, \citenamefont {Carvalho}, \citenamefont {Tueros},\ and\
  \citenamefont {Zas}}]{Alvarez_2012}%
  \BibitemOpen
  \bibfield  {author} {\bibinfo {author} {\bibfnamefont {J.}~\bibnamefont
  {Alvarez-Muñiz}}, \bibinfo {author} {\bibfnamefont {W.~R.}\ \bibnamefont
  {Carvalho}}, \bibinfo {author} {\bibfnamefont {M.}~\bibnamefont {Tueros}},\
  and\ \bibinfo {author} {\bibfnamefont {E.}~\bibnamefont {Zas}},\ }\bibfield
  {title} {\bibinfo {title} {{Coherent Cherenkov radio pulses from hadronic
  showers up to EeV energies}},\ }\href
  {https://doi.org/10.1016/j.astropartphys.2011.10.002} {\bibfield  {journal}
  {\bibinfo  {journal} {Astroparticle Physics}\ }\textbf {\bibinfo {volume}
  {35}},\ \bibinfo {pages} {287–299} (\bibinfo {year} {2012})}\BibitemShut
  {NoStop}%
\bibitem [{\citenamefont {Javaid}(2012)}]{Javaid2012}%
  \BibitemOpen
  \bibfield  {author} {\bibinfo {author} {\bibfnamefont {A.}~\bibnamefont
  {Javaid}},\ }\bibfield  {title} {\bibinfo {title} {{Monte Carlo simulation
  for radio detection of ultra-high energy air shower cores by ANITA-II}},\
  }\href@noop {} {\bibfield  {journal} {\bibinfo  {journal} {University of
  Delaware, Delaware (USA)}\ } (\bibinfo {year} {2012})}\BibitemShut {NoStop}%
\bibitem [{\citenamefont {Saftoiu}\ \emph {et~al.}(2013)\citenamefont
  {Saftoiu}, \citenamefont {Sima}, \citenamefont {Rebel}, \citenamefont
  {Badescu}, \citenamefont {Brancus}, \citenamefont {Haungs}, \citenamefont
  {Lazanu}, \citenamefont {Mitrica}, \citenamefont {Stanca},\ and\
  \citenamefont {Toma}}]{SAFTOIU_2013}%
  \BibitemOpen
  \bibfield  {author} {\bibinfo {author} {\bibfnamefont {A.}~\bibnamefont
  {Saftoiu}}, \bibinfo {author} {\bibfnamefont {O.}~\bibnamefont {Sima}},
  \bibinfo {author} {\bibfnamefont {H.}~\bibnamefont {Rebel}}, \bibinfo
  {author} {\bibfnamefont {A.}~\bibnamefont {Badescu}}, \bibinfo {author}
  {\bibfnamefont {I.}~\bibnamefont {Brancus}}, \bibinfo {author} {\bibfnamefont
  {A.}~\bibnamefont {Haungs}}, \bibinfo {author} {\bibfnamefont
  {I.}~\bibnamefont {Lazanu}}, \bibinfo {author} {\bibfnamefont
  {B.}~\bibnamefont {Mitrica}}, \bibinfo {author} {\bibfnamefont
  {D.}~\bibnamefont {Stanca}},\ and\ \bibinfo {author} {\bibfnamefont
  {G.}~\bibnamefont {Toma}},\ }\bibfield  {title} {\bibinfo {title} {Studies of
  radio emission from neutrino induced showers generated in rock salt},\ }\href
  {https://doi.org/https://doi.org/10.1016/j.astropartphys.2013.04.002}
  {\bibfield  {journal} {\bibinfo  {journal} {Astroparticle Physics}\ }\textbf
  {\bibinfo {volume} {46}},\ \bibinfo {pages} {1} (\bibinfo {year}
  {2013})}\BibitemShut {NoStop}%
\bibitem [{\citenamefont {de~Vries}\ \emph {et~al.}(2016)\citenamefont
  {de~Vries} \emph {et~al.}}]{deVries2016}%
  \BibitemOpen
  \bibfield  {author} {\bibinfo {author} {\bibfnamefont {K.~D.}\ \bibnamefont
  {de~Vries}} \emph {et~al.},\ }\bibfield  {title} {\bibinfo {title} {{The
  cosmic-ray air-shower signal in Askaryan radio detectors}},\ }\href@noop {}
  {\bibfield  {journal} {\bibinfo  {journal} {Astropart.Phys.}\ }\textbf
  {\bibinfo {volume} {74}},\ \bibinfo {pages} {96} (\bibinfo {year} {2016})},\
  \Eprint {https://arxiv.org/abs/arXiv:1503.02808} {arXiv:1503.02808}
  \BibitemShut {NoStop}%
\bibitem [{\citenamefont {De~Kockere}\ \emph {et~al.}(2021)\citenamefont
  {De~Kockere}, \citenamefont {de~Vries},\ and\ \citenamefont {van
  Eijndhoven}}]{DeKockere2021}%
  \BibitemOpen
  \bibfield  {author} {\bibinfo {author} {\bibfnamefont {S.}~\bibnamefont
  {De~Kockere}}, \bibinfo {author} {\bibfnamefont {K.}~\bibnamefont
  {de~Vries}},\ and\ \bibinfo {author} {\bibfnamefont {N.}~\bibnamefont {van
  Eijndhoven}},\ }\bibfield  {title} {\bibinfo {title} {{Simulation of the
  propagation of CR air shower cores in ice}},\ }\href
  {https://doi.org/10.22323/1.395.1032} {\bibfield  {journal} {\bibinfo
  {journal} {PoS}\ }\textbf {\bibinfo {volume} {ICRC2021}},\ \bibinfo {pages}
  {1032} (\bibinfo {year} {2021})}\BibitemShut {NoStop}%
\bibitem [{\citenamefont {De~Kockere}\ \emph {et~al.}(2023)\citenamefont
  {De~Kockere}, \citenamefont {de~Vries}, \citenamefont {van Eijndhoven},\ and\
  \citenamefont {Latif}}]{DeKockere2023}%
  \BibitemOpen
  \bibfield  {author} {\bibinfo {author} {\bibfnamefont {S.}~\bibnamefont
  {De~Kockere}}, \bibinfo {author} {\bibfnamefont {K.}~\bibnamefont
  {de~Vries}}, \bibinfo {author} {\bibfnamefont {N.}~\bibnamefont {van
  Eijndhoven}},\ and\ \bibinfo {author} {\bibfnamefont {U.~A.}\ \bibnamefont
  {Latif}},\ }\bibfield  {title} {\bibinfo {title} {{Simulation of the
  propagation of cosmic ray air showers in ice}},\ }\href
  {https://doi.org/10.22323/1.424.0015} {\bibfield  {journal} {\bibinfo
  {journal} {PoS}\ }\textbf {\bibinfo {volume} {ARENA2022}},\ \bibinfo {pages}
  {015} (\bibinfo {year} {2023})}\BibitemShut {NoStop}%
\bibitem [{\citenamefont {Huege}\ \emph {et~al.}(2013)\citenamefont {Huege},
  \citenamefont {Ludwig},\ and\ \citenamefont {James}}]{MainCoreas}%
  \BibitemOpen
  \bibfield  {author} {\bibinfo {author} {\bibfnamefont {T.}~\bibnamefont
  {Huege}}, \bibinfo {author} {\bibfnamefont {M.}~\bibnamefont {Ludwig}},\ and\
  \bibinfo {author} {\bibfnamefont {C.~W.}\ \bibnamefont {James}},\ }\bibfield
  {title} {\bibinfo {title} {{Simulating radio emission from air showers with
  CoREAS}},\ }in\ \href {https://doi.org/10.1063/1.4807534} {\emph {\bibinfo
  {booktitle} {AIP Conference Proceedings}}}\ (\bibinfo  {publisher} {AIP
  Publishing},\ \bibinfo {address} {Melville (New York)},\ \bibinfo {year}
  {2013})\BibitemShut {NoStop}%
\bibitem [{\citenamefont {{Belov, K. and Mulrey, K. and Romero-Wolf, A. and
  Wissel, S.A. and Zilles, A. and Bechtol, K. and others}}(2016)}]{SLAC_T510}%
  \BibitemOpen
  \bibfield  {author} {\bibinfo {author} {\bibnamefont {{Belov, K. and Mulrey,
  K. and Romero-Wolf, A. and Wissel, S.A. and Zilles, A. and Bechtol, K. and
  others}}},\ }\bibfield  {title} {\bibinfo {title} {{Accelerator Measurements
  of Magnetically Induced Radio Emission from Particle Cascades with
  Applications to Cosmic-Ray Air Showers}},\ }\href
  {https://doi.org/10.1103/PhysRevLett.116.141103} {\bibfield  {journal}
  {\bibinfo  {journal} {Phys. Rev. L}\ }\textbf {\bibinfo {volume} {116}},\
  \bibinfo {eid} {141103} (\bibinfo {year} {2016})},\ \Eprint
  {https://arxiv.org/abs/1507.07296} {arXiv:1507.07296 [astro-ph.IM]}
  \BibitemShut {NoStop}%
\bibitem [{\citenamefont {{U. A. Latif \textit{et
  al}}}(2023{\natexlab{a}})}]{Latif2023}%
  \BibitemOpen
  \bibfield  {author} {\bibinfo {author} {\bibnamefont {{U. A. Latif \textit{et
  al}}}},\ }\bibfield  {title} {\bibinfo {title} {{Propagating air shower radio
  signals to in-ice antennas}},\ }\href {https://doi.org/10.22323/1.424.0016}
  {\bibfield  {journal} {\bibinfo  {journal} {PoS}\ }\textbf {\bibinfo {volume}
  {424}},\ \bibinfo {pages} {016} (\bibinfo {year}
  {2023}{\natexlab{a}})}\BibitemShut {NoStop}%
\bibitem [{\citenamefont {{U. A. Latif \textit{et
  al}}}(2023{\natexlab{b}})}]{Latif_ICRC23}%
  \BibitemOpen
  \bibfield  {author} {\bibinfo {author} {\bibnamefont {{U. A. Latif \textit{et
  al}}}},\ }\bibfield  {title} {\bibinfo {title} {{Simulation of radio signals
  from cosmic-ray cascades in air and ice as observed by in-ice Askaryan radio
  detectors}},\ }\href {https://doi.org/https://doi.org/10.22323/1.444.0346}
  {\bibfield  {journal} {\bibinfo  {journal} {PoS}\ }\textbf {\bibinfo {volume}
  {ICRC2023}},\ \bibinfo {pages} {346} (\bibinfo {year}
  {2023}{\natexlab{b}})}\BibitemShut {NoStop}%
\bibitem [{\citenamefont {Bechtol}\ \emph {et~al.}(2004)\citenamefont
  {Bechtol}, \citenamefont {Belov}, \citenamefont {Borch}, \citenamefont
  {Chen}, \citenamefont {Clem}, \citenamefont {Gorham} \emph
  {et~al.}}]{kravchenko_besson_meyers_2004}%
  \BibitemOpen
  \bibfield  {author} {\bibinfo {author} {\bibfnamefont {K.}~\bibnamefont
  {Bechtol}}, \bibinfo {author} {\bibfnamefont {K.}~\bibnamefont {Belov}},
  \bibinfo {author} {\bibfnamefont {K.}~\bibnamefont {Borch}}, \bibinfo
  {author} {\bibfnamefont {P.}~\bibnamefont {Chen}}, \bibinfo {author}
  {\bibfnamefont {J.}~\bibnamefont {Clem}}, \bibinfo {author} {\bibfnamefont
  {P.}~\bibnamefont {Gorham}}, \emph {et~al.},\ }\bibfield  {title} {\bibinfo
  {title} {{In situ index-of-refraction measurements of the South Polar firn
  with the RICE detector}},\ }\href
  {https://doi.org/10.3189/172756504781829800} {\bibfield  {journal} {\bibinfo
  {journal} {Journal of Glaciology}\ }\textbf {\bibinfo {volume} {50}},\
  \bibinfo {pages} {522–532} (\bibinfo {year} {2004})}\BibitemShut {NoStop}%
\bibitem [{\citenamefont {Bechtol}\ \emph {et~al.}(2022)\citenamefont
  {Bechtol}, \citenamefont {Belov}, \citenamefont {Borch}, \citenamefont
  {Chen}, \citenamefont {Clem}, \citenamefont {Gorham} \emph
  {et~al.}}]{Bechtol_2022}%
  \BibitemOpen
  \bibfield  {author} {\bibinfo {author} {\bibfnamefont {K.}~\bibnamefont
  {Bechtol}}, \bibinfo {author} {\bibfnamefont {K.}~\bibnamefont {Belov}},
  \bibinfo {author} {\bibfnamefont {K.}~\bibnamefont {Borch}}, \bibinfo
  {author} {\bibfnamefont {P.}~\bibnamefont {Chen}}, \bibinfo {author}
  {\bibfnamefont {J.}~\bibnamefont {Clem}}, \bibinfo {author} {\bibfnamefont
  {P.}~\bibnamefont {Gorham}}, \emph {et~al.},\ }\bibfield  {title} {\bibinfo
  {title} {{SLAC T-510 experiment for radio emission from particle showers:
  Detailed simulation study and interpretation}},\ }\href
  {https://doi.org/10.1103/physrevd.105.063025} {\bibfield  {journal} {\bibinfo
   {journal} {Physical Review D}\ }\textbf {\bibinfo {volume} {105}},\ \bibinfo
  {pages} {063025} (\bibinfo {year} {2022})}\BibitemShut {NoStop}%
\bibitem [{\citenamefont {Kelley}\ \emph {et~al.}(2017)\citenamefont {Kelley},
  \citenamefont {Lu}, \citenamefont {Seckel}, \citenamefont {Pan},\ and\
  \citenamefont {Besson}}]{Kelley2018}%
  \BibitemOpen
  \bibfield  {author} {\bibinfo {author} {\bibfnamefont {J.}~\bibnamefont
  {Kelley}}, \bibinfo {author} {\bibfnamefont {M.-Y.}\ \bibnamefont {Lu}},
  \bibinfo {author} {\bibfnamefont {D.}~\bibnamefont {Seckel}}, \bibinfo
  {author} {\bibfnamefont {Y.}~\bibnamefont {Pan}},\ and\ \bibinfo {author}
  {\bibfnamefont {D.~Z.}\ \bibnamefont {Besson}},\ }\bibfield  {title}
  {\bibinfo {title} {{Observation of two deep, distant (1.4, 4)km impulsive RF
  transmitters by the Askaryan Radio Array (ARA).}},\ }\href
  {https://doi.org/10.22323/1.301.1030} {\bibfield  {journal} {\bibinfo
  {journal} {PoS}\ }\textbf {\bibinfo {volume} {ICRC2017}},\ \bibinfo {pages}
  {1030} (\bibinfo {year} {2017})}\BibitemShut {NoStop}%
\bibitem [{\citenamefont {Latif}(2020)}]{Latif_thesis_2020}%
  \BibitemOpen
  \bibfield  {author} {\bibinfo {author} {\bibfnamefont {U.~A.}\ \bibnamefont
  {Latif}},\ }\emph {\bibinfo {title} {{Towards measurement of UHECR with the
  ARA experiment}}},\ \href@noop {} {Ph.D. thesis},\ \bibinfo  {school}
  {University of Kansas} (\bibinfo {year} {2020})\BibitemShut {NoStop}%
\bibitem [{\citenamefont {Latif}(2023)}]{Latif_IceRayTracing_2020}%
  \BibitemOpen
  \bibfield  {author} {\bibinfo {author} {\bibfnamefont {U.}~\bibnamefont
  {Latif}},\ }\href@noop {} {\bibinfo {title} {{IceRayTracing}}},\ \bibinfo
  {howpublished} {\url{https://github.com/uzairlatif90/IceRayTracing}}
  (\bibinfo {year} {2023})\BibitemShut {NoStop}%
\bibitem [{\citenamefont {Schl\"uter}\ \emph {et~al.}(2020)\citenamefont
  {Schl\"uter}, \citenamefont {Gottowik}, \citenamefont {Huege},\ and\
  \citenamefont {Rautenberg}}]{Schluter:2020tdz}%
  \BibitemOpen
  \bibfield  {author} {\bibinfo {author} {\bibfnamefont {F.}~\bibnamefont
  {Schl\"uter}}, \bibinfo {author} {\bibfnamefont {M.}~\bibnamefont
  {Gottowik}}, \bibinfo {author} {\bibfnamefont {T.}~\bibnamefont {Huege}},\
  and\ \bibinfo {author} {\bibfnamefont {J.}~\bibnamefont {Rautenberg}},\
  }\bibfield  {title} {\bibinfo {title} {{Refractive displacement of the
  radio-emission footprint of inclined air showers simulated with CoREAS}},\
  }\href {https://doi.org/10.1140/epjc/s10052-020-8216-z} {\bibfield  {journal}
  {\bibinfo  {journal} {Eur. Phys. J. C}\ }\textbf {\bibinfo {volume} {80}},\
  \bibinfo {pages} {643} (\bibinfo {year} {2020})},\ \Eprint
  {https://arxiv.org/abs/2005.06775} {arXiv:2005.06775 [astro-ph.IM]}
  \BibitemShut {NoStop}%
\bibitem [{\citenamefont {Van~den Broeck}\ \emph {et~al.}(2023)\citenamefont
  {Van~den Broeck}, \citenamefont {Buitink}, \citenamefont {de~Vries},
  \citenamefont {Huege},\ and\ \citenamefont {Latif}}]{VandenBroeck2023}%
  \BibitemOpen
  \bibfield  {author} {\bibinfo {author} {\bibfnamefont {D.}~\bibnamefont
  {Van~den Broeck}}, \bibinfo {author} {\bibfnamefont {S.}~\bibnamefont
  {Buitink}}, \bibinfo {author} {\bibfnamefont {K.}~\bibnamefont {de~Vries}},
  \bibinfo {author} {\bibfnamefont {T.}~\bibnamefont {Huege}},\ and\ \bibinfo
  {author} {\bibfnamefont {U.~A.}\ \bibnamefont {Latif}},\ }\bibfield  {title}
  {\bibinfo {title} {{Radio propagation in non-uniform media}},\ }\href
  {https://doi.org/10.22323/1.424.0020} {\bibfield  {journal} {\bibinfo
  {journal} {PoS}\ }\textbf {\bibinfo {volume} {ARENA2022}},\ \bibinfo {pages}
  {020} (\bibinfo {year} {2023})}\BibitemShut {NoStop}%
\bibitem [{\citenamefont {Hecht}(2002)}]{Hecht2002}%
  \BibitemOpen
  \bibfield  {author} {\bibinfo {author} {\bibfnamefont {E.}~\bibnamefont
  {Hecht}},\ }\href@noop {} {\emph {\bibinfo {title} {{Optics}}}}\ (\bibinfo
  {publisher} {Addison-Wesley},\ \bibinfo {year} {2002})\BibitemShut {NoStop}%
\bibitem [{\citenamefont {{C. Glaser \textit{et al}}}(2020)}]{NuRadioMC}%
  \BibitemOpen
  \bibfield  {author} {\bibinfo {author} {\bibnamefont {{C. Glaser \textit{et
  al}}}},\ }\bibfield  {title} {\bibinfo {title} {{NuRadioMC: simulating the
  radio emission of neutrinos from interaction to detector}},\ }\href
  {https://doi.org/10.1140/epjc/s10052-020-7612-8} {\bibfield  {journal}
  {\bibinfo  {journal} {European Physical Journal C}\ }\textbf {\bibinfo
  {volume} {80}},\ \bibinfo {eid} {77} (\bibinfo {year} {2020})}\BibitemShut
  {NoStop}%
\bibitem [{\citenamefont {Lehtinen}\ \emph {et~al.}(2004)\citenamefont
  {Lehtinen}, \citenamefont {Gorham}, \citenamefont {Jacobson},\ and\
  \citenamefont {Roussel-Dupr\'e}}]{Lehtinen_2004}%
  \BibitemOpen
  \bibfield  {author} {\bibinfo {author} {\bibfnamefont {N.~G.}\ \bibnamefont
  {Lehtinen}}, \bibinfo {author} {\bibfnamefont {P.~W.}\ \bibnamefont
  {Gorham}}, \bibinfo {author} {\bibfnamefont {A.~R.}\ \bibnamefont
  {Jacobson}},\ and\ \bibinfo {author} {\bibfnamefont {R.~A.}\ \bibnamefont
  {Roussel-Dupr\'e}},\ }\bibfield  {title} {\bibinfo {title} {{FORTE satellite
  constraints on ultrahigh energy cosmic particle fluxes}},\ }\href
  {https://doi.org/10.1103/physrevd.69.013008} {\bibfield  {journal} {\bibinfo
  {journal} {Physical Review D}\ }\textbf {\bibinfo {volume} {69}},\ \bibinfo
  {pages} {013008} (\bibinfo {year} {2004})}\BibitemShut {NoStop}%
\bibitem [{\citenamefont {Corstanje}\ \emph {et~al.}()\citenamefont
  {Corstanje}, \citenamefont {Buitink}, \citenamefont {Desmet}, \citenamefont
  {Falcke} \emph {et~al.}}]{Corstanje_Code}%
  \BibitemOpen
  \bibfield  {author} {\bibinfo {author} {\bibfnamefont {A.}~\bibnamefont
  {Corstanje}}, \bibinfo {author} {\bibfnamefont {S.}~\bibnamefont {Buitink}},
  \bibinfo {author} {\bibfnamefont {M.}~\bibnamefont {Desmet}}, \bibinfo
  {author} {\bibfnamefont {H.}~\bibnamefont {Falcke}}, \emph {et~al.},\
  }\bibfield  {title} {\bibinfo {title} {{A high-precision interpolation method
  for pulsed radio signals from cosmic-ray air showers}},\ }\href
  {https://doi.org/10.1088/1748-0221/18/09/p09005} {\bibfield  {journal}
  {\bibinfo  {journal} {Journal of Instrumentation}\ }\textbf {\bibinfo
  {volume} {18}}\bibinfo  {number} { (09)},\ \bibinfo {pages} {P09005
  (2023)}}\BibitemShut {NoStop}%
\bibitem [{\citenamefont {for Environmental
  Information~(NCEI)}()}]{magn_field}%
  \BibitemOpen
\bibfield  {number} {  }\bibfield  {author} {\bibinfo {author} {\bibfnamefont
  {N.~C.}\ \bibnamefont {for Environmental Information~(NCEI)}},\ }\href@noop
  {} {\bibinfo {title} {{Geomagnetic Calculators}}},\ \bibinfo {howpublished}
  {\url{https://www.ngdc.noaa.gov/geomag/calculators/magcalc.shtml\#igrfwmm}}\BibitemShut
  {NoStop}%
\bibitem [{\citenamefont {Heck}\ and\ \citenamefont {Knapp}(1998)}]{Heck1998}%
  \BibitemOpen
  \bibfield  {author} {\bibinfo {author} {\bibfnamefont {D.}~\bibnamefont
  {Heck}}\ and\ \bibinfo {author} {\bibfnamefont {J.}~\bibnamefont {Knapp}},\
  }\bibfield  {title} {\bibinfo {title} {{Upgrade of the Monte Carlo Code
  CORSIKA to Simulate Extensive Air Showers with Energies $> 10^{20}$ eV}},\
  }\href {https://publikationen.bibliothek.kit.edu/270043705} {\bibfield
  {journal} {\bibinfo  {journal} {Report}\ }\textbf {\bibinfo {volume} {FZKA
  6097}} (\bibinfo {year} {1998})}\BibitemShut {NoStop}%
\bibitem [{\citenamefont {Kobal}(2001)}]{KOBAL2001259}%
  \BibitemOpen
  \bibfield  {author} {\bibinfo {author} {\bibfnamefont {M.}~\bibnamefont
  {Kobal}},\ }\bibfield  {title} {\bibinfo {title} {{A thinning method using
  weight limitation for air-shower simulations}},\ }\href
  {https://doi.org/https://doi.org/10.1016/S0927-6505(00)00158-4} {\bibfield
  {journal} {\bibinfo  {journal} {Astroparticle Physics}\ }\textbf {\bibinfo
  {volume} {15}},\ \bibinfo {pages} {259} (\bibinfo {year} {2001})}\BibitemShut
  {NoStop}%
\bibitem [{\citenamefont {Ostapchenko}(2011)}]{QGSJET}%
  \BibitemOpen
  \bibfield  {author} {\bibinfo {author} {\bibfnamefont {S.}~\bibnamefont
  {Ostapchenko}},\ }\bibfield  {title} {\bibinfo {title} {{Monte Carlo
  treatment of hadronic interactions in enhanced Pomeron scheme: QGSJET-II
  model}},\ }\href {https://doi.org/10.1103/PhysRevD.83.014018} {\bibfield
  {journal} {\bibinfo  {journal} {Phys. Rev. D}\ }\textbf {\bibinfo {volume}
  {83}},\ \bibinfo {pages} {014018} (\bibinfo {year} {2011})}\BibitemShut
  {NoStop}%
\bibitem [{\citenamefont {Bass}(1998)}]{UrQMD_1}%
  \BibitemOpen
  \bibfield  {author} {\bibinfo {author} {\bibfnamefont {S.}~\bibnamefont
  {Bass}},\ }\bibfield  {title} {\bibinfo {title} {{Microscopic models for
  ultrarelativistic heavy ion collisions}},\ }\href
  {https://doi.org/10.1016/s0146-6410(98)00058-1} {\bibfield  {journal}
  {\bibinfo  {journal} {Progress in Particle and Nuclear Physics}\ }\textbf
  {\bibinfo {volume} {41}},\ \bibinfo {pages} {255–369} (\bibinfo {year}
  {1998})}\BibitemShut {NoStop}%
\bibitem [{\citenamefont {Bleicher}\ \emph {et~al.}(1999)\citenamefont
  {Bleicher}, \citenamefont {Zabrodin}, \citenamefont {Spieles}, \citenamefont
  {Bass}, \citenamefont {Ernst}, \citenamefont {Soff}, \citenamefont {Bravina},
  \citenamefont {Belkacem}, \citenamefont {Weber}, \citenamefont {Stöcker},\
  and\ \citenamefont {Greiner}}]{UrQMD_2}%
  \BibitemOpen
  \bibfield  {author} {\bibinfo {author} {\bibfnamefont {M.}~\bibnamefont
  {Bleicher}}, \bibinfo {author} {\bibfnamefont {E.}~\bibnamefont {Zabrodin}},
  \bibinfo {author} {\bibfnamefont {C.}~\bibnamefont {Spieles}}, \bibinfo
  {author} {\bibfnamefont {S.~A.}\ \bibnamefont {Bass}}, \bibinfo {author}
  {\bibfnamefont {C.}~\bibnamefont {Ernst}}, \bibinfo {author} {\bibfnamefont
  {S.}~\bibnamefont {Soff}}, \bibinfo {author} {\bibfnamefont {L.}~\bibnamefont
  {Bravina}}, \bibinfo {author} {\bibfnamefont {M.}~\bibnamefont {Belkacem}},
  \bibinfo {author} {\bibfnamefont {H.}~\bibnamefont {Weber}}, \bibinfo
  {author} {\bibfnamefont {H.}~\bibnamefont {Stöcker}},\ and\ \bibinfo
  {author} {\bibfnamefont {W.}~\bibnamefont {Greiner}},\ }\bibfield  {title}
  {\bibinfo {title} {{Relativistic hadron-hadron collisions in the
  ultra-relativistic quantum molecular dynamics model}},\ }\href
  {https://doi.org/10.1088/0954-3899/25/9/308} {\bibfield  {journal} {\bibinfo
  {journal} {J. Phys. G: Nucl. Part. Phys.}\ }\textbf {\bibinfo {volume}
  {25}},\ \bibinfo {pages} {1859} (\bibinfo {year} {1999})}\BibitemShut
  {NoStop}%
\bibitem [{\citenamefont {de~Vries}\ \emph {et~al.}(2010)\citenamefont
  {de~Vries}, \citenamefont {van~den Berg}, \citenamefont {Scholten},\ and\
  \citenamefont {Werner}}]{de_Vries_2010}%
  \BibitemOpen
  \bibfield  {author} {\bibinfo {author} {\bibfnamefont {K.~D.}\ \bibnamefont
  {de~Vries}}, \bibinfo {author} {\bibfnamefont {A.~M.}\ \bibnamefont {van~den
  Berg}}, \bibinfo {author} {\bibfnamefont {O.}~\bibnamefont {Scholten}},\ and\
  \bibinfo {author} {\bibfnamefont {K.}~\bibnamefont {Werner}},\ }\bibfield
  {title} {\bibinfo {title} {{The lateral distribution function of coherent
  radio emission from extensive air showers: Determining the chemical
  composition of cosmic rays}},\ }\href
  {https://doi.org/10.1016/j.astropartphys.2010.08.003} {\bibfield  {journal}
  {\bibinfo  {journal} {Astroparticle Physics}\ }\textbf {\bibinfo {volume}
  {34}},\ \bibinfo {pages} {267–273} (\bibinfo {year} {2010})}\BibitemShut
  {NoStop}%
\bibitem [{\citenamefont {Schröder}(2017)}]{Schr_der_2017}%
  \BibitemOpen
  \bibfield  {author} {\bibinfo {author} {\bibfnamefont {F.~G.}\ \bibnamefont
  {Schröder}},\ }\bibfield  {title} {\bibinfo {title} {{Radio detection of
  cosmic-ray air showers and high-energy neutrinos}},\ }\href
  {https://doi.org/10.1016/j.ppnp.2016.12.002} {\bibfield  {journal} {\bibinfo
  {journal} {Prog. Part. Nucl. Phys.}\ }\textbf {\bibinfo {volume} {93}},\
  \bibinfo {pages} {1–68} (\bibinfo {year} {2017})}\BibitemShut {NoStop}%
\bibitem [{\citenamefont {Scholten}\ \emph {et~al.}(2016)\citenamefont
  {Scholten}, \citenamefont {Trinh}, \citenamefont {Bonardi}, \citenamefont
  {Buitink}, \citenamefont {Correa}, \citenamefont {Corstanje} \emph
  {et~al.}}]{Scholten_2016}%
  \BibitemOpen
  \bibfield  {author} {\bibinfo {author} {\bibfnamefont {O.}~\bibnamefont
  {Scholten}}, \bibinfo {author} {\bibfnamefont {T.~N.~G.}\ \bibnamefont
  {Trinh}}, \bibinfo {author} {\bibfnamefont {A.}~\bibnamefont {Bonardi}},
  \bibinfo {author} {\bibfnamefont {S.}~\bibnamefont {Buitink}}, \bibinfo
  {author} {\bibfnamefont {P.}~\bibnamefont {Correa}}, \bibinfo {author}
  {\bibfnamefont {A.}~\bibnamefont {Corstanje}}, \emph {et~al.},\ }\bibfield
  {title} {\bibinfo {title} {Measurement of the circular polarization in radio
  emission from extensive air showers confirms emission mechanisms},\ }\href
  {https://doi.org/10.1103/physrevd.94.103010} {\bibfield  {journal} {\bibinfo
  {journal} {Physical Review D}\ }\textbf {\bibinfo {volume} {94}},\ \bibinfo
  {pages} {103010} (\bibinfo {year} {2016})}\BibitemShut {NoStop}%
\bibitem [{\citenamefont {Alameddine}\ \emph
  {et~al.}(2023{\natexlab{a}})\citenamefont {Alameddine} \emph
  {et~al.}}]{CORSIKA8_1}%
  \BibitemOpen
  \bibfield  {author} {\bibinfo {author} {\bibfnamefont {J.-M.}\ \bibnamefont
  {Alameddine}} \emph {et~al.} (\bibinfo {collaboration} {CORSIKA}),\
  }\bibfield  {title} {\bibinfo {title} {{The particle-shower simulation code
  CORSIKA 8}},\ }\href {https://doi.org/10.22323/1.444.0310} {\bibfield
  {journal} {\bibinfo  {journal} {PoS}\ }\textbf {\bibinfo {volume}
  {ICRC2023}},\ \bibinfo {pages} {310} (\bibinfo {year}
  {2023}{\natexlab{a}})},\ \Eprint {https://arxiv.org/abs/2308.05475}
  {arXiv:2308.05475 [astro-ph.HE]} \BibitemShut {NoStop}%
\bibitem [{\citenamefont {Alameddine}\ \emph
  {et~al.}(2023{\natexlab{b}})\citenamefont {Alameddine} \emph
  {et~al.}}]{CORSIKA8_2}%
  \BibitemOpen
  \bibfield  {author} {\bibinfo {author} {\bibfnamefont {J.-M.}\ \bibnamefont
  {Alameddine}} \emph {et~al.} (\bibinfo {collaboration} {CORSIKA}),\
  }\bibfield  {title} {\bibinfo {title} {{Simulations of cross media showers
  with CORSIKA 8}},\ }\href {https://doi.org/10.22323/1.444.0442} {\bibfield
  {journal} {\bibinfo  {journal} {PoS}\ }\textbf {\bibinfo {volume}
  {ICRC2023}},\ \bibinfo {pages} {442} (\bibinfo {year}
  {2023}{\natexlab{b}})},\ \Eprint {https://arxiv.org/abs/2309.05897}
  {arXiv:2309.05897 [astro-ph.IM]} \BibitemShut {NoStop}%
\bibitem [{\citenamefont {Heck}\ and\ \citenamefont
  {Pierog}()}]{CORSIKA_manual}%
  \BibitemOpen
  \bibfield  {author} {\bibinfo {author} {\bibfnamefont {D.}~\bibnamefont
  {Heck}}\ and\ \bibinfo {author} {\bibfnamefont {T.}~\bibnamefont {Pierog}},\
  }\href@noop {} {\emph {\bibinfo {title} {{Extensive Air Shower Simulation
  with CORSIKA: A User’s Guide (Version 7.7500 from April 14, 2023)}}}},\
  \bibinfo {organization} {Karlsruher Institut f\"ur Technologie (KIT)},\
  \bibinfo {note} {available at
  \url{https://www.iap.kit.edu/corsika/70.php}}\BibitemShut {NoStop}%
\end{thebibliography}%

\end{document}